\documentclass[prb,aps, superscriptaddress, twocolumn]{revtex4-1}
\usepackage{amsmath,amssymb}
\usepackage{graphicx}
\usepackage{wasysym}
\usepackage{amsfonts}
\usepackage{bm}
\usepackage{enumerate}
\usepackage{color}
\usepackage[resetlabels]{multibib}
\usepackage{epstopdf}
\usepackage{latexsym}
\usepackage[breaklinks,colorlinks = true,linkcolor = red,urlcolor=cyan,citecolor=red]{hyperref}

\usepackage{times}
\newcommand{\bea}{\begin{eqnarray}}
\newcommand{\eea}{\end{eqnarray}}
\newcommand{\be}{\begin{eqnarray}}
\newcommand{\ee}{\end{eqnarray}}
\newcommand{\bw}{\begin{widetext}}
\newcommand{\ew}{\end{widetext}}
\newcommand{\nn}{\nonumber}
\newcommand{\bs}{\boldsymbol}
\newcommand{\la}{\langle}
\newcommand{\ra}{\rangle}

\newcommand{\bsbr}{{\bar{ \boldsymbol {r}}}}

\def\ket#1{{|#1\rangle}}

\def\br{\vecsymb{r}}

\def\btau{{\bm \tau}}
\def\vecsymb#1{\boldsymbol{#1}}
\def\nss{\mathcal{S}}
\def\va{\vecsymb{a}}

\def\vr{\vecsymb{r}}

\def\vk{\vecsymb{k}}

\begin{document}
\title{Fractionalizing glide reflections in two-dimensional $Z_2$ topologically ordered phases}
\author{SungBin Lee}
\email{sungbinl@uci.edu}
\affiliation{Department of Physics, Korea Advanced Institute of Science and Technology, Daejeon 305-701, Korea}
\affiliation{Department of Physics and Astronomy, University of California, Irvine, CA 92697, USA}
\author{Michael Hermele}
\email{Michael.Hermele@colorado.edu}
\affiliation{Department of Physics, 390 UCB, University of Colorado, Boulder, CO 80309, USA}
\affiliation{Center for Theory of Quantum Matter, University of Colorado, Boulder, CO 80309, USA}

\author{S. A. Parameswaran}
\email{sidp@uci.edu}
\affiliation{Department of Physics and Astronomy, University of California, Irvine, CA 92697, USA}

\date{\today}
\begin{abstract}
We study the fractionalization of space group symmetries in two-dimensional topologically ordered phases. Specifically, we focus on $Z_2$-fractionalized phases in two dimensions whose deconfined topological excitations transform trivially under translational symmetries, but projectively under {glide reflections}, whose quantum numbers are hence fractionalized. We accomplish this by generalizing the dichotomy between even and odd gauge theories to incorporate additional symmetries inherent to {non-symmorphic} crystals. We show that the resulting fractionalization of point group quantum numbers can be detected in numerical studies of ground state wave functions.
We illustrate these ideas using a microscopic model of a system of bosons at integer unit cell filling on a lattice with space group $p4g$, that can be mapped to a half-magnetization plateau for an $S=1/2$ spin system on the Shastry-Sutherland lattice. 
\end{abstract}
\maketitle
\section{Introduction}
\label{sec:intro}

What are the different patterns of quantum number fractionalization that can emerge at long wavelengths and low energies in condensed matter systems? Answering this question has taken on added urgency following the compelling numerical demonstration that a gapped, fractionalized quantum spin liquid phase emerges as the ground state of the Heisenberg model on the frustrated two-dimensional kagome lattice.\cite{yan2011spin} This observation has sparked renewed activity in studying detailed properties of fractionalized phases, with the goal of identifying additional distinguishing features that may aid the detection of their characteristic {topological order} in real materials or engineered systems.  
 
The line of enquiry we pursue in this paper is to examine the role of symmetry in topologically ordered phases. Indeed, though not required for topological order --- which can be defined  in terms of fractionalization of particle {statistics} --- one or more symmetries must exist for us to have quantum numbers to fractionalize! Such symmetry-enriched topologically ordered phases are  richer in their phenomenology  --- intuitively, detecting fractional {\it statistics} requires a non-local measurement, {\it e.g.} interferometry, while fractional {\it charges} may be observed by more conventional probes, such as spectroscopy, transport, or tunneling.~\cite{WenPSG,PhysRevB.87.104406,PhysRevB.87.155115, HungWen}

We focus specifically on  
 the interplay of the fractionalization of two quantum numbers: namely, the conserved charge corresponding to a global $U(1)$ symmetry, and the set of quantum numbers of discrete space-group symmetries. Examples of systems with these symmetries include lattice models of bosons or fermions with a conserved charge or of spins with a conserved magnetization. 
The intimate link between the existence of these symmetries and fractionalization was made by Oshikawa and Hastings, building on earlier work by Lieb, Schulz, and Mattis in one dimension and is embodied in the {\it Hastings-Oshikawa-Lieb-Schulz-Mattis} (HOLSM) theorem.\cite{Lieb1961407,oshikawa2000topological,PhysRevB.69.104431,hastings2005sufficient} 
In a crystal with a fractional $U(1)$ charge per unit cell (hereafter termed the {\it filling}, and denoted $\nu$), the only 
known gapped insulating ground states
consistent with the HOLSM theorem either fractionalize the $U(1)$ charge, or else break translational symmetries. 

Most early work on the role of the HOLSM theorem in insulators
considered the fractionalization of only {\it one} of the two symmetries necessary to apply the theorem: namely, the on-site `internal'  $U(1)$ charge conservation symmetry. Charge fractionalization, while striking, is relatively intuitive: we simply `split' the charge-carrying particle into fractional pieces, a picture that has received dramatic experimental vindication in the context of the fractional quantum Hall effect. More recently, it has been recognized that in some insulators at fractional filling, $U(1)$ charge fractionalization  
is necessarily accompanied by the fractionalization of crystal momentum --- {\it i.e.}, the quantum number of lattice translations.\cite{paramekanti2004extending,essin2014spectroscopic,StationQSurfaceHOLSM} 
This holds in particular for fractionalized insulators
with $Z_2$ topological order, on which we focus in this work.
These states have one type of bosonic quasiparticle carrying $U(1)$ charge of 1/2 and
$Z_2$ gauge charge  --- termed a ``spinon'' or ``chargon'' depending on the context ---  and another --- the ``vison''--- which is charge-neutral and carries
the $Z_2$ gauge flux.  At fractional filling, visons necessarily carry fractional crystal momentum, as we discuss in detail below.

For integer $\nu$, however, the original HOLSM arguments 
{based on} translational symmetries are silent on the nature of  symmetry-preserving gapped  phases. {The applicability of symmetry-based considerations may be significantly enhanced, however,  upon inclusion of additional {\it point group} symmetries. Such symmetries extend the HOLSM arguments in cases when the point group structure is embedded with the translational symmetries in such a manner that the resulting space group is {\it non-symmorphic}  --- i.e., lacks a single point left invariant by all point group operations, modulo lattice translations. \cite{parameswaran2013topological}
Upon considering the full space group, it is possible to show that in order for a gapped phase to preserve  the full set of spatial symmetries without triggering fractionalization, the filling $\nu$ must be an integer multiple of an integer $\mathcal{S}$, dubbed the non-symmorphic rank, an invariant that characterizes a particular space group. When $\mathcal{S}=1$, evidently there are no additional constraints beyond those imposed by translational symmetry, and we must be satisfied with the original HOLSM theorems. However, when $\mathcal{S}>1$ (as it is for all the 157 non-symmorphic space groups), we see that absence of symmetry breaking is evidence of fractionalization even at integer fillings, thereby extending the HOLSM arguments to these new cases.}

Whereas both integer and fractional $\nu$ imply topological order and charge fractionalization in symmetric non-symmorphic insulators, they result in different fractionalization of space-group symmetries. For the $Z_2$ fractionalized phases we study here (with charge-neutral visons and half-charged chargons/spinons)  crystal momentum fractionalization is possible only at fractional filling; when $\nu$ is an integer, the fractionalized quasiparticles transform linearly, rather than projectively, under translations.\cite{paramekanti2004extending} However, as we show in this paper, they {\it do} transform projectively under a subset of the point-group symmetries that includes at least one non-symmorphic operation. Therefore, the models studied here represent a distinct space-group fractionalization class compared to those previously studied: while the fractionalization of translation symmetry alone is trivial, there is non-trivial fractionalization of point-group symmetry operations.  

{We note that  non-symmorphicity is by no means an exotic space-group property: over two-thirds of the 230 three-dimensional space groups have $\mathcal{S}>1$, including such mineralogically ubiquitous ones as the diamond ($Fd\bar{3}m$) and hexagonal close-packed ($P6_3/mmc$) structures. However, in this paper we focus on the simpler case of two dimensions, where of the 17 distinct {\it wallpaper} groups (as 2D space groups are termed) $4$ are non-symmorphic by virtue of including one or more {\it glide reflections}: a reflection in a plane followed by translation in the plane of reflection by one-half of a lattice vector. (This is the only possible non-symmorphic operation\footnote{In $d=3$, the large number of non-symmorphic space groups stems from the additional possibility of {\it screw rotations}, a rotation combined with a fractional lattice translation parallel to the rotation axis.} in $d=2$.) Of these, we further restrict ourselves to the maximally symmetric non-symmorphic 2D space group, termed $p4g$, as it will serve to illustrate our ideas and is relevant to several materials. Results for the other three cases should follow straightforwardly by relaxing one or more of the symmetries considered here.   

A subsidiary motivation to focus on $d=2$ is that understanding the interplay of topological order and symmetry in $d$ dimensions  is a powerful route to understanding and classifying {\it symmetry-protected topological phases} (SPT phases) in $d+1$ dimensions. The  only possible symmetry-preserving gapped surface theory of a $d+1$ dimensional SPT phase is a topologically ordered phase where the symmetries are implemented in a manner that is impossible in a 
strictly $d$-dimensional theory. Recently, SPT phases involving non-symmorphic crystalline symmmetries ---  so-called topological non-symmorphic crystalline insulators (TNCIs) --- have been proposed~\cite{TNCI_PRB_Original,FangFu,HourglassFermions,CohomologicalInsulators,TNCI_Classification}. We anticipate that the results reported here, in combination with the `flux-fusion anomaly test'\cite{HermeleChenFFAnomaly}, will assist in identifying interacting TNCIs.

{Here, we study a specific example of  symmetry fractionalization in the $p4g$ space group. This is the space group that characterizes the so-called {\it Shastry-Sutherland} lattice familiar in studies of frustrated magnetism; this lattice encodes $p4g$ symmetry by arranging $s$-orbitals in a unit cell at positions related via a glide symmetry.\cite{shastry1981exact} Materials realizing this structure include 
SrCu$_2$(BO$_3$)$_2$, Yb$_2$Pt$_2$Pb, TmB$_4$ and ErB$_4$.\cite{sebastian2008fractalization,PhysRevB.77.144425,michimura2006magnetic,1742-6596-51-1-011} 
In the context of spin systems, our analysis will bear on studies of  half-magnetization plateaus in an applied field.\cite{jaime2012magnetostriction,michimura2006magnetic,1742-6596-51-1-011, 1742-6596-200-3-032041,matsuda2013magnetization} (In an appendix, we also discuss a model that encodes $p4g$ symmetry in a different fashion, through non-trivial orbital quantum numbers associated with distinct sites, and obtain similar results.)  We note that while previous studies have introduced general frameworks for considering symmetries in $2+1$ dimensional topological phases~\cite{PhysRevB.87.104406,StationQ_Gcrossed} and have analyzed fractionalization of translational symmetry by visons~\cite{QiChengFangVisons, PhysRevB.87.104406,essin2014spectroscopic}, to our knowledge the interplay of  fractionalization with non-symmorphic symmetries has not been studied previously.

The remainder of this paper is organized as follows. 
In Sec.~~\ref{sec:overview}, we first provide a pedagogical overview of HOLSM theorems and their connection to odd Ising gauge theory, in the relatively familiar context of topologically ordered phases at $\nu=1/2$ on the square lattice. 
Through this discussion, the obstruction to applying such theorems at integer $\nu$, and the crucial role of non-symmorphicity in extending them to this case, will become apparent. 
{We also provide a review of space-group symmetry fractionalization}; readers familiar with the HOLSM theorem and symmetry fractionalization can skip this pedagogical review and proceed directly to {Sec.~~\ref{sec:models}, where we introduce 
a simple lattice model that will be the focus of our study. We construct an effective gauge theory describing fractionalization on this lattice in Sec.~~\ref{sec:Z4-gauge} 
and demonstrate the manner in which glide symmetry is fractionalized. 
 Finally, in Sec.~\ref{sec:detection} we discuss how to detect the fractionalization via studies of the entanglement structure of ground state wave functions, and identify which aspects of the symmetry fractionalization are in principle accessible to experimental probes.  We also include, in an appendix, a discussion of a model with exactly the same space group but with a different lattice representation of the symmetries, and show that the universal aspects of symmetry fractionalization remain unchanged.}

\section{Commensurability, Fractionalization, and Space Group Symmetry: Overview}
\label{sec:overview}

We begin by providing a brief review of the commensurability conditions required by the HOLSM theorems, and elaborate on the interplay between the fractionalization of the $U(1)$ and spatial symmetries. 

\subsection{HOLSM Theorems}
\label{subsec:HOLSM}

Let us first specify the setting in which the HOLSM commensurability theorems may be applied. We study lattice systems with a specified space group $\mathcal{G}$,  described by a local Hamiltonian $\hat{H}$ that preserves all the symmetries of $\mathcal{G}$. In addition, we assume that the system has (at least) one global charge $\hat{\mathcal{Q}}$ conserved by $\hat{H}$, {\it i.e.} $[H,\hat{\mathcal{Q}}]=0$ (we may thus replace $\hat{\mathcal{Q}}$ by its $c$-number expectation value $\mathcal{Q}$ throughout). We make no assumptions as to the origins of the conserved charge, so for instance the systems we consider could be built out of (i) spinless fermions or bosons, where $\mathcal{Q}$ is just the conserved particle number; (ii) spinful fermions with SU(2) spin symmetry, in which case $\mathcal{Q}$ is one-half the total fermion number (since the two spin components may be treated independently); or (iii) lattice spins with (at least) U(1) spin rotational invariance, in which case we may take 
the charge on lattice site $\br$ to be $S+ \hat{S}^z_{\br}$  where $S, \hat{S}^z_{\br}$ are its total spin and magnetization, and define $\mathcal{Q}$ accordingly. Considering a finite system with $\mathcal{N}_c = N_1\times N_2\times N_3 $ unit cells (of course $N_3=1$ for the $d=2$ case), we may then define the {\it filling} to be the charge per unit cell, {\it i.e.} $\nu =  \mathcal{Q}/\mathcal{N}_c$, which will be held fixed in the thermodynamic limit.
 We impose periodic boundary conditions that identify  ${\br}$ and $ {\br}+N_i\va_i$. Note that in the case when $\nu$ is a fraction, we choose $N_c$ so to ensure that $\mathcal{Q} = \nu N_c$ is an integer, in accord with the quantization of charge. We work in units where $\hbar=e=1$, so that the quantum of flux is $2\pi$.

 The original HOLSM theorem states that at fractional $\nu$, it is impossible for the system to have a unique, translationally-invariant gapped ground state. This leaves open the following possibilities for an insulating ground state:
\begin{enumerate}[(i)]
\item {\bf The system remains gapless.}  
\item {\bf The system is gapped, and breaks translational symmetry}, thereby enlarging the unit cell. The effective filling in the new unit cell is then an integer.  
\item {\bf The system is gapped, and preserves translational symmetries.} In this case, we have a ground state degeneracy that cannot be associated with a broken symmetry. One route to this is for the low-energy excitations to be fractionalized, so that the emergent low-energy description is the deconfined phase of a lattice gauge theory\cite{kogut1979introduction,rothe2012lattice}; the degenerate ground states may then be associated with processes that create a quasiparticle-hole pairs from the vacuum, and thread them around a noncontractible loop (here we assume periodic boundary conditions) before fusing them back into the vacuum. The resulting states are topologically distinct from the original ground state, but no local observable   can tell them apart.
\end{enumerate}
We exclusively focus on case (iii) in this paper. 

We now sketch the argument that leads to these conclusions; we give an intuitive proof, and refer the reader to Refs.~\onlinecite{Lieb1961407,oshikawa2000topological,PhysRevB.69.104431, hastings2005sufficient, altman1998haldane, oshikawa1997magnetization} for a more formal treatment. We begin with a ground state $|\Psi\rangle$ and thread a flux quantum through a periodic direction, which, by gauge invariance, returns us to the original Hamiltonian. This procedure produces an eigenstate $|\tilde{\Psi}\rangle$. Earlier work has argued that for an insulator, $|\tilde{\Psi}\rangle$ must be a `low energy' state, {\it i.e.} its energy approaches that of the ground state in the thermodynamic limit.\cite{oshikawa2003insulator, paramekanti2004extending, PhysRevB.69.104431, hastings2005sufficient} 
 \footnote{Although rigorous energy bounds can only be given for a different -- but gauge-equivalent -- flux insertion, for pedagogical reasons we keep a simpler choice with the understanding that the arguments of Ref.~\onlinecite{hastings2005sufficient} can be applied to make our proofs rigorous.}
The key step is to show that $|\tilde{\Psi}\rangle$ is distinct from  $|\Psi\rangle$, which would then establish ground state degeneracy. In the case of fractional filling, these states differ in crystal momentum\cite{oshikawa2003insulator,paramekanti2004extending,PhysRevB.69.104431, hastings2005sufficient}, as we now demonstrate. 

For specificity\footnote{We can generalize the results to any lattice and to $d=3$, but for the cases studied in this paper, $d=2$ and square lattice symmetry is sufficient}, let us take the example of a square lattice in $d=2$ with lattice spacing $a$. Now, imagine adiabatically threading $2\pi$ flux through a handle of the torus (recall that our system is defined with periodic boundary conditions), say the one enclosed by a noncontractible loop parallel to the $x$-axis. The particles are assumed to couple minimally to this gauge flux, with unit charge. Suppose we began with a ground state $\ket{\Psi}$; as translation in the $x$-direction is a symmetry, we may assume that this is a state of fixed momentum: in other words, we have
\be
\hat{T}_x \ket{\Psi} = e^{i P_0} \ket{\Psi}
\ee for some crystal momentum $P_0$.  After inserting $2\pi$ flux, we assume that we are in a new state $\ket{\Psi'}$; however, it is clear that the flux insertion can be implemented via a uniform vector potential $\boldsymbol{A} = \frac{2\pi}{N_xa}  \hat{\mathbf{x}}$ and that this preserves $T_x$ as a symmetry. Thus, we may conclude that $\hat{T}_x \ket{\Psi'} = e^{i P_0} \ket{\Psi'}$, i.e. the crystal momentum is unchanged during the adiabatic flux threading. However, under the flux insertion, the Hamiltonian also changes, from $\hat{H}(0)$ to $\hat{H}(2\pi)$ --- it now describes a system with an inserted flux. 

We complete the adiabatic cycle by performing a large gauge transformation, implemented by the operator
\be
\hat{U} = \exp\left\{i\frac{2\pi}{N_x} \int d^2 r\,   \hat{\mathbf{x}} \cdot\br  \hat\rho(\br)\right\}
\ee
where we have employed second-quantized notation and  $\hat{\rho}(\br)$ is the density operator corresponding to the conserved charge $\hat{\mathcal Q}$. It is straightforward to show that
\be
\hat{T}_x \hat{U}\hat{T}_x^{-1} \hat{U}^{-1} =  e^{i\frac{2\pi}{N_x} \int \hat{\rho}(\br)} = e^{ i 2\pi \mathcal{Q}/N_x} = e^{i 2\pi \nu N_y}.
\ee
Using this, we see that the final result of the adiabatic cycle is a state $\ket{\tilde{\Psi}} = \hat{U} \ket{\Psi'}$, whose crystal momentum $P_1$ is given by
\be
e^{iP_1}\ket{\tilde{\Psi}} \equiv \hat{T}_x \ket{\tilde\Psi} &=& \hat{T}_x \hat{U}\ket{\Psi'} \\
 &=& e^{i 2\pi \nu N_y}  \hat{U}\hat{T}_x \ket{\Psi'}  
= e^{i (P_0 + 2\pi \nu N_y)}\ket{\tilde\Psi}. \nonumber
\ee
Clearly, if $N_y$ is chosen relatively prime to $q$ (note that we may still choose $N_x$ so that $\nu N_x N_y$ is an integer, so there is no inconsistency) then $\ket{\Psi}$, $\ket{\tilde{\Psi}}$ differ in their crystal momentum, {\it i.e.} the ground state after flux insertion is distinct from the one we began with. Given that $\ket{\tilde{\Psi}}$ is a low-energy state degenerate with the ground state in the thermodynamic limit, we have a ground-state degeneracy, and therefore  the system must fall into one of the three categories discussed above.

\subsubsection{Extending HOLSM to Integer Filling}
If we attempt to apply the above arguments at integer filling ($\nu \in Z$), it is clear that the change in momentum upon flux insertion is always a reciprocal lattice vector: in other words, we cannot use crystal momentum to  differentiate between $|\Psi\rangle$ and $|\tilde{\Psi}\rangle$. However, on non-symmorphic lattices, one can still distinguish these states using the quantum numbers of the non-symmorphic operations\cite{parameswaran2013topological}. Let us review how this argument proceeds. For simplicity, since we are working at integer $\nu$, we may take $N_i = N$. Now, consider a non-symmorphic symmetry ${G}$ that involves a point-group transformation $g$ followed by a translation through a fraction of a lattice vector $\btau$ in a direction left invariant by $g$: in other words, we have
\be
G: \br \rightarrow g\br +\btau.
\ee
In this paper, we will be concerned with the case when $g$ is a mirror reflection, in which case $\tau$ is always one-half a lattice vector, and ${G}$ is termed a {\it glide reflection}. This is the only possible non-symmorphic symmetry in $d=2$.

As before, we begin with a ground state $\ket{\Psi}$, and assume it is an eigenstate of  
${G}$, {\it i.e.}
\be
\hat{G}\ket{\Psi} = e^{i \theta} \ket{\Psi}
\ee
We consider the smallest reciprocal lattice vector $\vk$    left invariant by $g$, so that $g\vk = \vk$ and $\vk$ generates the invariant sublattice along $\hat{\vk}$.   We now thread flux by introducing a vector potential $\boldsymbol{A} = \vk/N$. (Note that as $\vk$ is in the reciprocal lattice, $\vk \cdot \va_i$  is always an integer multiple of $2\pi$, so this is always a pure gauge flux; the case studied above is simply a specific instance of this.) In the process of flux insertion $\ket{\Psi}$ evolves to a state $\ket{\Psi'}$ that is degenerate with it. Once again, to compare $\ket{\Psi'}$ to $\ket{\Psi}$, we must return to the original gauge, which can be accomplished by the unitary transformation $\ket{\Psi'} \rightarrow \hat{U}_{\vk}\ket{\Psi'}\equiv|\tilde{\Psi}\rangle $,  where
\be
\hat{U}_{\vk} = \exp\left\{{ \frac{i}{N} \int d^d{r}\, \vk\cdot\vr \hat{\rho}(\br) }\right\}
\ee
removes the  inserted flux. 
Since  $\boldsymbol{A}$ is left invariant
 by  ${G}$, threading flux does not alter $\hat{G}$ eigenvalues, so $\ket{\Psi}$ and $\ket{\Psi'}$ have the same quantum number under $\hat{G}$; however, on acting with $\hat{U}_{\vk}$, the eigenvalue changes, as can be computed from the equation:
\be\label{eq:FCR}
\hat{G} \hat{U}_{\vk}\hat{G}^{-1} =  \hat{U}_{\vk} e^{2\pi i \Phi_g(\vk){\mathcal{Q}}/N}\ee
where  we have defined the phase factor $ \Phi_g(\vk) = \btau\cdot \vk/{2\pi}$, and ${\mathcal Q}=\nu N^2$ is the total charge on the $N\times N$ torus.
 It may be readily verified that since $g\vk=\vk$, $\Phi_g(\vk)$ is unchanged by a shift in real-space origin. For a non-symmorphic symmetry operation ${G}$,  this phase  $\Phi_g(\vk)$  must be a fraction. This follows since $\btau$ is a fractional translation. (If a lattice translation had the same projection onto $\vk$ as $\btau$,  this would yield an integer phase factor.\cite{konig1997electronic} However, this would render the screw/glide removable  {\it i.e.} reduced to point group element$\times$translation by change of origin and hence not truly non-symmorphic.)  Thus, for  $\hat{G}$  non-symmorphic, $\Phi_g(\vk) = p/\nss_G$, with $p,\nss_G$ relatively prime.
From (\ref{eq:FCR}) we conclude that $\ket{\Psi}$ and $|\tilde{\Psi}\rangle$
 have distinct $\hat{G}$ eigenvalues whenever $\Phi_g(\vk) \mathcal{Q}/N  =p  N\nu/\nss_G $ is a fraction. Since we may always choose $N$ relatively prime to the $\nss_G$, the result of flux insertion is a state distinguished from the original state by its $\hat{G}$ eigenvalue, unless the filling is a multiple of $\nss_G$. For a glide $\nss_G=2$. 
 
From this argument, we see that in any 2D crystal with a glide reflection plane, we can extend the applicability of the HOLSM theorem to {\it odd integer} fillings, by considering ground states that are invariant under the glide symmetry (in addition to translations). Similar arguments can be made also for screw rotations in $d=3$ (and a modified approach can be constructed for two `exceptional' 3D space groups where this approach fails\cite{parameswaran2013topological}), but we focus on the $d=2$ case in this paper. Note that different arguments that do not invoke flux insertion can also be made~\cite{Watanabe24112015} but we will not discuss these here.

\subsection{Topological Order, Gauge Theories and Crystal Momentum Fractionalization}
\label{subsec:odd-ising-gauge}
As we have discussed, assuming the absence of symmetry breaking and the presence of a gap, the HOLSM theorems require a ground-state degeneracy on the torus, and that the ground states differ by crystal momenta or other point group symmetry quantum numbers. For the square lattice at half-filling, an effective low-energy description that is consistent with this picture is that the ground state exhibits ${Z}_2$ topological order~\cite{balents02,balents99,kitaev03,moessner01a,moessner01b,read91,chakraborty89,senthil00,wen91,SenthilFisher}
. This is a fractionalized, translationally invariant insulating phase, whose ground state is not unique in a multiply connected geometry ({\it e.g.}, the periodic boundary condition-torus considered here) owing to the presence of a gapped $Z_2$ vortex or {\it vison} excitation in the spectrum. The degeneracy is then associated with the presence or absence of a vison threading a non-contractible loop of the torus and is hence topological.  
The splitting between the vison/no vison states vanishes exponentially with system size in the thermodynamic limit, since the tunneling of a vison `into' or `out of' the torus costs an energy that scales with $L$, as the vison is a gapped bulk excitation.

With these preliminaries, we are ready to study the fractionalization of symmetries in our 
square lattice example, assuming $\nu$ is an integer or half-integer. We introduce the effective low-energy theory for the topological phase:  introducing Ising  degrees of freedom  $\tau^\mu_{\bs r \bs r'}$ ($\mu = x,y,z$) on each link $(\bs r \bs r')$ on the square lattice, we have the Ising gauge theory Hamiltonian
\bea
\mathcal{H}_{\rm IGT} = - h \sum_{\la \bs r \bs r' \ra } \tau^x_{\bs r \bs r'} 
-K \sum_{p } \underset{ \bs r \bs r' \in p }{\prod} \tau^z_{\bs r \bs r'},
\label{eq:HIGTSquare}
\eea
with a Gauss law constraint for every site $\bs r$ 
\be
 \underset{ \bs r' \in  \la \bs r \bs r' \ra  }{\prod} \tau^x_{\bs r \bs r'} = (-1)^{2\nu},
 \label{eq:GaussLawIGTSquare}
\ee
where $\la \cdots \ra$ labels nearest-neighbor sites and and $p$ labels plaquettes. 
As the microscopic origins of effective gauge theories such as $\mathcal{H}_{\rm IGT}$ are detailed in several excellent references~\cite{balents02,balents99,kitaev03,moessner01a,moessner01b,read91,chakraborty89,senthil00,wen91,SenthilFisher}, and since we also give a detailed account of similar constructions in the non-symmorphic case below, we do not repeat them here. Note that
$\mathcal{H}_{\rm IGT}$ is in an ordered phase for $K \ll h$ and is in a deconfined phase for $K \gg h$. It is the latter that has a finite-energy gapped vison excitation and hence has topological order. As shown in Ref.~\onlinecite{paramekanti2004extending}, linking the Gauss law constraint to the filling through (\ref{eq:GaussLawIGTSquare}) is required in order to satisfy the commensurability constraints of the HOLSM theorem: it guarantees that the topologically degenerate ground states differ by the appropriate crystal momentum computed via the flux insertion argument.

In the $K\gg h$ deconfined phase there is a gapped {\it chargon} (or spinon, in the case when the $U(1)$ is a component of spin) that carries fractional electromagnetic charge (or fractional spin quantum number) $\nu e$ as well as a  $Z_2$ Ising charge. It also supports the dual excitation to a chargon, a gapped Ising vortex or ``vison''  that carries  $Z_2$ gauge flux; the vison is neutral under the global $U(1)$ charge. Note that we have made a specific choice of $Z_2$ fractionalization in taking the chargon/spinon to carry fractional global $U(1)$ charge and the vison to be neutral; we will focus exclusively on this choice in throughout this paper. 
 We have assumed the vison is a boson, and will also take the chargon to be a boson.
As one carries $Z_2$ charge and the other $Z_2$ flux, they acquire a mutual statistics phase of $\pi$, rendering their bound state a fermion. (Note that the chargon/spinon has been formally integrated out and is not a dynamical excitation of $\mathcal{H}_{\text{IGT}}$, which describes a pure gauge theory.)

For the purposes of understanding crystal momentum fractionalization, the central point to note is that when $\nu$ is a half-odd integer, the Gauss law (\ref{eq:GaussLawIGTSquare}) requires a non-trivial static background Ising gauge charge  on each site --- in other words, there is a chargon frozen at each site.   
In this manner, the HOLSM  commensurability constraints force us to work with an {\it odd} Ising gauge theory in the terminology of Ref.~\onlinecite{moessner01b}.  As we now review, this requires that the vison carry a non-zero crystal momentum. 

A single-vison state is one where a single plaquette $p$ has a non-zero $Z_2$ magnetic flux, {\it i.e.} \be \underset{ \bs r \bs r' \in p }{\prod} \tau^z_{\bs r \bs r'} = -1.\label{eq:visonplaqdef} \ee
This is  non-local in the $\tau^z$-basis of the gauge theory: to satisfy (\ref{eq:visonplaqdef}), we must flip  $\tau^z$s on a string of bonds out to infinity. In order to further elucidate symmetry properties of the vison, it is convenient to move to a description in which the vison creation operator is local. This is accomplished by mapping our $Z_2$ gauge theory into its transverse-field Ising model (TFIM) dual, as we now detail. 

Let us define dual lattice sites $\bsbr$ that reside at the center of each square lattice plaquette.  
We define a new set of Ising operators $\sigma^\mu_{\bsbr}$ for the TFIM;  these are related to the original Ising variables via
\bea
\tau_{\bs r \bs r'}^x &=& \eta_{\bsbr \bsbr'} \sigma_{\bsbr }^z \sigma_{\bsbr'}^z
\label{eq:10} \\
\underset{ \bs r \bs r' \in p_{\bsbr} }{\prod} \tau^z_{\bs r \bs r'}
&=& \sigma_{\bsbr}^x , 
\label{eq:11}
\eea
where $\bsbr$ and $\bsbr'$ are the dual lattice sites on either side of the the link $(\bs r \bs r' )$ in the original square lattice, and $p_{\bsbr}$ represents the direct lattice plaquette centered on dual lattice site $\br$. The $Z_2$ phase $\eta_{\bsbr \bsbr'}$ is defined on each link $(\bsbr \bsbr')$ of the dual lattice. 
Under this mapping, the IGT Hamiltonian \eqref{eq:HIGTSquare} is transformed into the Hamiltonian for the TFIM
\bea
\mathcal{H}_{\rm TFIM} = - h \sum_{\bsbr \bsbr '} \eta_{\bsbr \bsbr '} \sigma_{\bsbr}^z \sigma_{\bsbr '}^z - K \sum_{\bsbr} \sigma_{\bsbr}^x
\label{eq:HTFIMSquare}
\eea
while the Gauss law \eqref{eq:GaussLawIGTSquare} appears as a
 phase constraint
\bea
\underset{ \bsbr \bsbr ' \in p }{\prod} \eta_{\bsbr \bsbr '} = (-1)^{2\nu}.
\label{eq:TFImFrustrationSquare}
\eea
 As a final step to obtain a well-defined TFIM, we must pick a prescription for the phases $\eta_{\bsbr \bsbr '}$ that satisfies (\ref{eq:TFImFrustrationSquare}); while for integer $\nu$ we may simply choose $\eta_{\bsbr\bsbr'} =1$, for half-odd integer $\nu$ this `gauge fixing' amounts to choosing one {\it frustrated} bond on each plaquette, where the Ising coupling flips sign. The distinction between even and odd Ising gauge theory (integer and half-odd integer $\nu$) is thus mapped into the distinction between ordinary and ({fully}) frustrated TFIMs~\cite{moessner01b}. Upon gauge-fixing, the vison becomes a local operator --- essentially $\sigma_{\bsbr}^z$ with an appropriate (gauge-fixed) phase --- and therefore we can consider the symmetry properties of vison states by considering the transformation properties of $\sigma_{\bsbr}^z$. 
 For the case of odd Ising gauge theory, any gauge fixing satisfying (\ref{eq:TFImFrustrationSquare}) apparently breaks translation symmetry (in other words, there is \emph{no} pattern of frustrated bonds that respects the translational symmetry of the lattice).  Translation symmetry is in fact still present, but must act on $\sigma^z_{\bar{\boldsymbol{r}}}$ combined with appropriate gauge transformations.  As a result, the action of translation on any single vison state satisfies
  \be
T_{\bs x}^v T_{\bs y }^v\left( T_{\bs x}^{v}\right)^{-1} \left( T_{\bs y}^{v}\right)^{-1} = -1,
\label{eq:visonPSGtrans}
\ee
where $T_{\bs x}^v$ and $T_{\bs y}^v$ are the translations of visons by lattice vectors ${\bs x}$ and ${\bs y}$ on a square lattice. 

In other words, visons transform {\it projectively} under the translations when $\nu$ is a half-odd-integer. (Such non-commuting translations should be familiar from the case of a charged particle in a magnetic field, but that situation is distinct from the example considered here.) 

An intuitive understanding of why HOLSM requires the vison to carry nontrivial crystal momentum is as follows. Imagine starting from the deconfined phase of $\mathcal{H}_{\text{IGT}}$  and tuning parameters so that we condense one of its gapped excitations. Recall that the vison and spinon are both bosons, but have nontrivial mutual statistics --- so either can be condensed, but this will confine the other.  Now, as the chargon carries a global $U(1)$ charge, it follows that condensing this will break the $U(1)$ symmetry, leading to a superfluid with gapless excitations. In contrast, condensing the vison  leads to a gapped, confined phase. Were the vison to carry trivial quantum numbers, then the resulting vison-condensed phase would be a gapped phase without any fractionalization or translational symmetry breaking, {\it which is impossible for fractional $\nu$}. Thus, consistency demands that the vison carry non-zero crystal momentum, as this will trigger broken symmetry when it condenses.

The discussion above has all the basic ingredients that will permeate the remainder of this paper: a commensurability condition (HOLSM theorem) requiring a certain non-trivial background gauge charge in the low-energy effective theory, in turn forcing a fractionalization of crystal symmetry quantum numbers. However, the alert reader will note that we have only considered the fractionalization of translational quantum numbers.  
Are there distinct signatures of the fractionalization of other crystal symmetries?  In particular, we ask: are there situations where visons transform regularly under translations, so that the projective symmetry arises {\it purely} as a property of the other crystal symmetries?

The answer to this is in the affirmative, and perhaps unsurprisingly, is intimately connected to the extension of the HOLSM theorem to integer filling discussed above. In the balance of this paper, we detail how to generalize the `odd filling $\leftrightarrow$ odd Ising gauge theory' connection to integer fillings in non-symmorphic 2D crystals by constructing low-energy effective descriptions of fractionalized phases in these crystals. We then examine the signatures of fractionalization of glide reflection symmetry in numerics.

\section{Model}
\label{sec:models}
We now begin the central analysis of this paper, where we explore space group symmetry fractionalization at integer filling. We will consider the case of the 2D non-symmorphic space group $p4g$, which has $\mathcal{S} =2$, at filling $\nu=1$. This is precisely a situation where the only commensurability condition is the extended HOLSM theorem that relies on the glide reflection symmetries in $p4g$. We note that  among the 17 2D space groups (also termed {\it wallpaper} groups) there are three other non-symmorphic groups: $pg, p2mg$ and $p2gg$. However, these have reduced symmetry compared to $p4g$, and are apparently of less relevance to materials. While we do not consider these in detail, we expect that the broad features of glide symmetry fractionalization should apply to these as well.
We restrict our attention to a (generalized) Bose-Hubbard model.  Denoting the boson creation and annihilation operators at site $\bs r$ by $B_{\bs r}^\dagger$ and $B_{\bs r}$, we have 
\be
{\mathcal H} = -t \sum_{\la \bs r \bs r' \ra} (B_{\bs r}^\dagger B_{\bs r'} + \text{h.c.}) + V[ \{ N_{\bs r} \} ],
\label{eq:bosonHamBase}
\ee
where the first term hops bosons between nearest-neighbors $\la \bs r \bs r' \ra$ and the second term is an interaction that depends on the boson number on site $\br$, $N_{\bs r} = B^\dagger_{\bs r} B_{\bs r}$. 
For $t \gg V$, the boson kinetic energy dominates and we enter a superfluid phase with broken $U(1)$ symmetry. We are primarily interested in the opposite limit, $t \ll V$, where the system is in a Mott insulating phase. In particular, we focus on the case  $\nu=1$, 
where the extended HOLSM theorem requires that any symmetry-preserving insulator exhibit fractionalization. 

 Our central example consists of a model of $s$-orbitals arranged on the sites of the so-called Shastry-Sutherland lattice (SSL), familiar in the context of frustrated magnetism; here, the glide symmetry operates by interchanging the spatial coordinates of the four orbitals in a unit cell 
 (Fig.~\ref{fig:SSLbase}), but the orbitals themselves transform trivially. At $\nu=1$, as there are four orbitals in each unit cell, we have an average site filling of $1/4$. 
 
 While this $s$-orbital boson model could be realized and studied in cold atom systems, we note that it also has significant relevance to frustrated magnetism of spins on the Shastry-Sutherland lattice. This follows from taking the hard core limit of (\ref{eq:bosonHamBase}); in this limit, by associating the presence (absence) of a boson on a site with spin up (down) with respect to a reference axis, we arrive at a Hamiltonian for $U(1)$ spins, with the total charge related to the total magnetization along the specified axis. Such a spin Hamiltonian can describe the magnetization process of spins in a field, with the Mott insulating phases corresponding to magnetization plateaus.

 We note that there are a variety of  Shastry-Sutherland lattice materials that exhibit magnetization plateaus, such as SrCu$_2$(BO$_3$)$_2$, Yb$_2$Pt$_2$Pb, TmB$_4$ and ErB$_4$. Of these, the plateau at $1/2$ the saturation magnetization has been observed in SrCu$_2$(BO$_3$)$_2$, TmB$_4$ and ErB$_4$.\cite{kageyama1999exact,jaime2012magnetostriction,michimura2006magnetic,1742-6596-51-1-011} We will present  a detailed numerical study of the structure of the $\nu=1/2$ phase diagram elsewhere.
 
In Appendix~~\ref{app:dorbitalglidefrac} we also examine a different model, containing two orthogonally oriented $d$-orbitals in each unit cell. This shares the same $p4g$ space group as the Shastry-Sutherland lattice, but encodes the symmetries in its nontrivial orbital content, rather than in terms of the spatial coordinates of the orbitals within a unit cell. While we relegate detailed discussion of this model to the appendix, the symmetry fractionalization is essentially identical to that for the Shastry-Sutherland example; the sole technical difference is that there is a  straightforward parton construction that directly leads to a $\mathbb{Z}_2$ gauge theory, rather than accessing this via the intermiedate step of a $\mathbb{Z}_4$ theory. Nevertheless, the $d$-orbital model serves as a check that our results depend only on the space group rather than a particular implementation of the symmetry in our model.

 \section{Glide Fractionalization} 
\label{sec:Z4-gauge}
As noted above, we will  focus on the example of the Shastry-Sutherland $s$-orbital model; readers interested in seeing how the story plays out in the $d$-orbital model are referred to Appendix ~~\ref{app:dorbitalglidefrac}.

\subsection{$Z_4$ Gauge Theory}
We begin our study of the $s$-orbital SSL model by constructing a parton effective theory that appropriately captures the fractionalization of the degrees of freedom. Recall that we are working at $\nu=1$ and there are four sites in each unit cell; therefore, an intuitive choice for a translationally-invariant parton mean-field solution is to split each of the `bare' charge-$1$ bosons  into four partons each with $1/4$ charge, and engineer a translationally-invariant parton Mott insulator in which each parton is frozen on a single site. Operationally, this may be implemented by introducing a rotor representation of these $1/4$-charges: we define a parton number operator $n_{\br}$, with conjugate phase variable $\phi_{\br}$. In terms of these operators, we may rewrite our original theory in terms of parton variables via the mapping
\be
B^\dagger_{\br} &=& (b^\dagger_{\br})^4 \label{eq:partonz4map}, \\
n_{\br} &=& 4 N_{\br} \label{eq:partonz4constraint}
\ee
where  $b^\dagger_{\br} = e^{i\phi_{\br}}$ is the parton creation operator. Eq. (\ref{eq:partonz4constraint}) should be viewed as a constraint on the parton Hilbert space that reduces it to the physical Hilbert space of the original bosons. 

To proceed, we follow the standard procedure of softening the constraint: we enlarge the Hilbert space to include parton configurations not satisfying (\ref{eq:partonz4constraint}), but compensate for this by introducing a gauge field whose role is to implement the constraint by dynamically projecting out unphysical degrees of freedom; a minimal choice for the gauge group is $Z_4$. Recall that gauge fields are associated with links of the lattice. We take the Hilbert space on a single link $\ell =(\br,\br')$ to consist of four states, $\mathfrak{h}_\ell = \{\ket{0},\ket{1},\ket{2},\ket{3}\}$, and introduce  a $Z_4$ vector potential operator $a_\ell$ and the associated  electric field operator $e_{\ell}$ via their action on $\mathfrak{h}_\ell$:
\bea
a_{\bs r \bs r'} \ket{k} = e^{ 2\pi i k/4} \ket{k} \,\,\,\text{   and   }\,\,\,
e_{\bs r \bs r'} \ket{k} =  \ket{k+1}, \label{eq:linkactionZ4}
\eea
where we identify $\ket{4}\equiv \ket{0}$. 
As links may be traversed in either direction, given a fiducial orientation used to define (\ref{eq:linkactionZ4}), for the reversed orientation we have
\bea
a_{\bs r' \bs r} = a_{\bs r \bs r'}^\dagger \,\,\,\text{ and} \,\,\,
e_{\bs r' \bs r} = e_{\bs r \bs r'}^\dagger.
\label{eq:reversedlinkZ4vars}
\eea
In terms of these variables,  the constraint (\ref{eq:partonz4constraint}) appears as a Gauss law for the $Z_4$ electric field sourced by the partons:
 \be
\underset{\bs r ' \in \la \bs r \bs r' \ra}{\prod} e_{\bs r \bs r'} = e^{ 2\pi i n_{\bs r} /4},
\label{eq:z4gausslaw}
\ee
where the product is over the links connecting $\br$ to its neighboring sites  $\br'$. 

The above considerations allow us to rewrite (\ref{eq:bosonHamBase}) as an effective gauge theory,
\bea
{{\mathcal H}} &=&  {{\mathcal H}_g} +{{\mathcal H}_m}\nn\\
 {{\mathcal H}_m} &=& -t \sum_{\la \bs r \bs r' \ra} ( b^\dagger_{\bs r} a_{\bs r \bs r'} b_{\bs r'} + \text{h.c.} ) + \tilde{V} [ \{ n_{\bs r} \} ],\label{eq:z4gaugeHam}
\\
{{\mathcal H}_g} &=& - h \sum_{\la \bs r \bs r' \ra} ( e_{\bs r \bs r'} + \text{h.c.}) - K \sum_{p\in \Box, \vartriangle} (\underset{ \bs r \bs r'\in p}{  \prod} a_{\bs r \bs r'} + \text{h.c.} ) \nn  \eea
where $h,t,K>0$ and the sum over $p$  ranges over the two different types of lattice plaquettes (triangular and square plaquettes) as shown in Fig.~\ref{fig:SSLbase}, and we have assumed $K_\Box = K_\vartriangle = K$. 
The link product for each plaquette assumes a fixed orientation, here taken to be anticlockwise. $\mathcal{H}_g$ represents the dynamics of the gauge field, required in order to implement  (\ref{eq:partonz4constraint}).  $\mathcal{H}_m$ represents the parton degrees of freedom that interact with the gauge field via the usual minimal coupling on the lattice, and $\tilde{V} $ is a parton interaction term written as  a function of the parton density $n_{\bs r}$.  We must specify how lattice symmetries act on the new fields $e_{\bs r \bs r'}, a_{\bs r \bs r'}$. Consider a symmetry operation $S$ such that $S:{\bs r\rightarrow s(\bs r)}$; we then require that the fields transform via 
\bea\label{eq:fieldtransformations}
S: e_{\bs r \bs r'}&\rightarrow& e_{s(\bs r) s(\bs r')},\nonumber\\
S: a_{\bs r \bs r'}&\rightarrow& a_{s(\bs r) s(\bs r')},
\eea
{\it i.e.}, $S$ acts on  $e_{\bs r \bs r'}, a_{\bs r \bs r'}$ simply by its action on their labels.

\begin{figure}
\includegraphics[width=0.5\columnwidth]{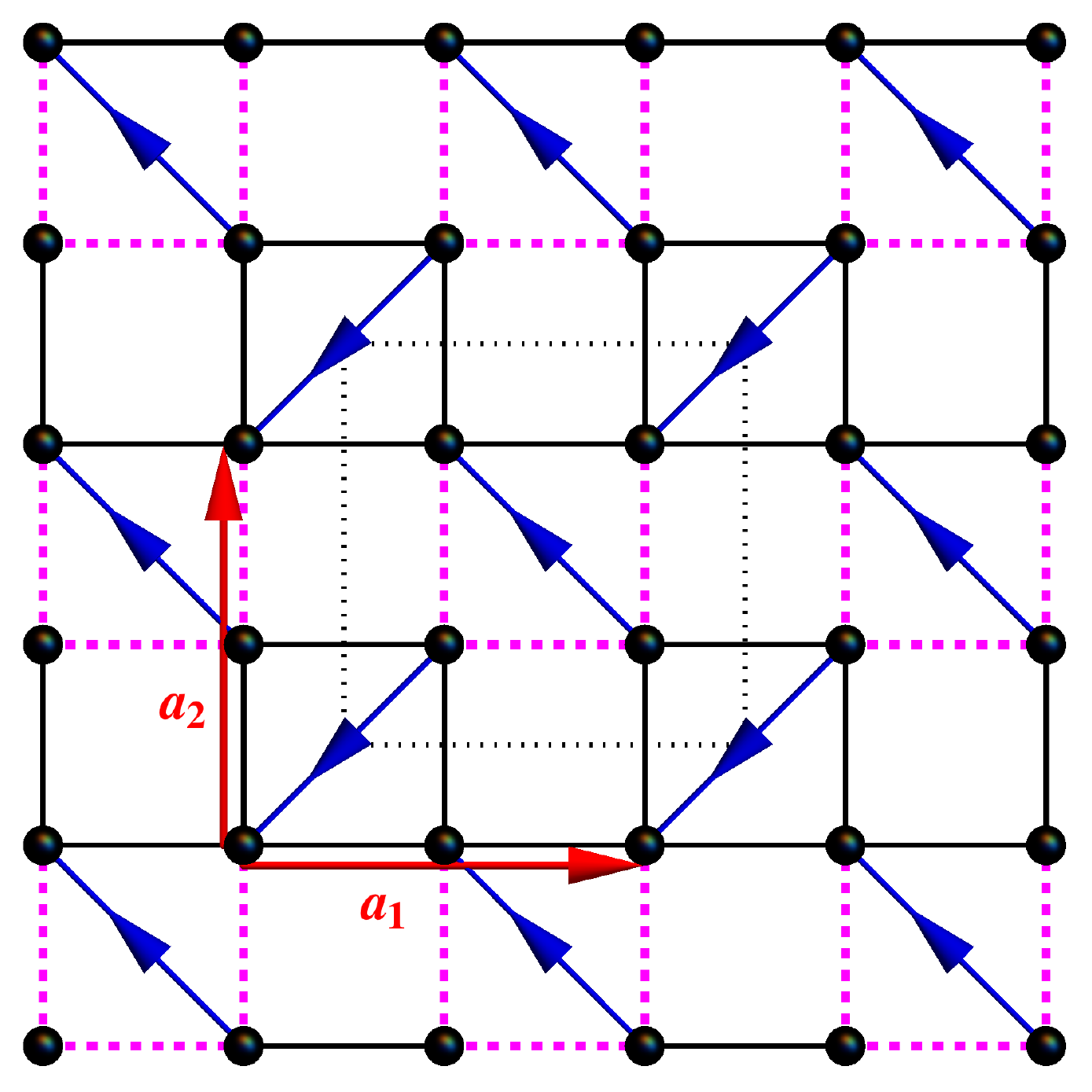}
\caption{(color online) {\bf SSL structure and $Z_4$ electric field configuration at $\nu=1$ on the SSL.} Lattice vectors $\bs{a}_1$ and $\bs{a}_2$. The  Gauss law constraint $\underset{\bs r ' \in \la \bs r \bs r' \ra}{\prod} e_{\bs r \bs r'} = i$ must be satisfied at each site. Solid (Black) and dashed (magenta) links indicate electric field $e_{\bs r \bs r'} = 1$ and -1 respectively. Blue arrows on the diagonal links represent $e_{\bs r \bs r'} = i$ when the link is traversed in the direction of the arrow. The dotted square shows a single unit cell that includes four sites; note that the $e$-field pattern  is the same in every unit cell, but breaks point-group symmetries.}
\label{fig:SSLbase}
\end{figure}

For $h\gg K$, the low-energy configurations of (\ref{eq:z4gaugeHam}) have $e_{\br\br'} =1$, and  from the Gauss law (\ref{eq:z4gausslaw}) we conclude that in this parameter regime $n_{\bs r } \equiv 0\,(\text{mod } 4)$. This is the confining phase of the gauge theory: the partons remain tied together into the original bosons, and therefore there are no states in the Hilbert space where free charges can be asymptotically separated. 

We are more interested in the limit of $K\gg h$, where the partons may move independently of each other and the electric field is no longer confined. We may consider the parton fields to be gapped, and assume that the interaction term $\tilde{V}$ is such that the partons form  a strong Mott insulating ground state with $n_{\bs r} =1$; since we have a single parton on each site, it is evident that this phase is translationally invariant. As the partons are gapped, we may integrate them out; the resulting theory is a pure $Z_4$ gauge theory (that will have the same form as $\mathcal{H}_g$, possibly with renormalized parameters). For low-energy configurations below the parton gap, we have $n_{\bs r} =1$ and therefore we may rewrite the Gauss law (\ref{eq:z4gausslaw}) as
\bea
\underset{\bs r ' \in \la \bs r \bs r' \ra}{\prod} e_{\bs r \bs r'}  = i. \label{eq:frozenz4gauss}
\eea
Observe that (\ref{eq:frozenz4gauss}) requires a non-trivial $Z_4$ gauge charge on every site; this is the $Z_4$ analog of the nontrivial background charge in the odd Ising gauge theory. 
From our discussion above, it is straightforward to see that for filling $\nu$ of the original bosons, the RHS is $e^{2\pi i \nu/4}$.
We show an electric field configuration satisfying (\ref{eq:frozenz4gauss}) in Fig.~~\ref{fig:SSLbase}. Note that, unlike in the case of the odd Ising gauge theory, the $Z_4$ electric field configuration does not enlarge the unit cell --- reflecting the preservation of translational symmetries --- but breaks point group symmetries, specifically the glide symmetry. We will explore the consequences of this shortly. In contrast, for boson filling $\nu=2$,  it is straightforward to see that all the space group symmetries are preserved by field configurations that satisfy the Gauss law, as shown in Fig.~~\ref{fig:SSLEfieldNu2}. 

The lowest-energy excitations of the pure gauge theory are {\it visons}\footnote{Note that the term `vison' is conventionally associated with Ising gauge theories; we use the term here more generally to describe fluxes in discrete gauge theories.}  that are constructed by inserting  $Z_4$ flux $\underset{\bs r \bs r'}{\prod} {a_{\bs r \bs r'}}$ on a single plaquette (of either shape).  
Since the electric field operator shifts the value of ${a_{\bs r \bs r'}}$, and we wish to only change the flux through a single plaquette, it follows that in order to create a vison we must apply electric field operators along a `string' of bonds on the lattice. Fig.~~\ref{fig:FluxInsSSL} shows an example of such a flux insertion operator $\hat{F}_\otimes$ at the square plaquette labeled $\otimes$:
\be
\hat{F}_\otimes = \underset{\bs r \bs r'  \rightarrow \infty }{\prod} e^{(\dagger)}_{\bs r \bs r'} = e_{ 1  2} e^\dagger_{ 2  3} e_{ 3  4} e^\dagger_{ 4  5} \cdots ,
\label{eq:visonstringoperatorZ4}
\ee
where the indices $1,2,3 \cdots $ label each site along the `string' identified in Fig.~~\ref{fig:FluxInsSSL}. 
In order to further examine the vison properties, it is once again convenient to move to a dual representation of the $Z_4$ gauge theory.


\begin{figure}
\includegraphics[width=0.5\columnwidth]{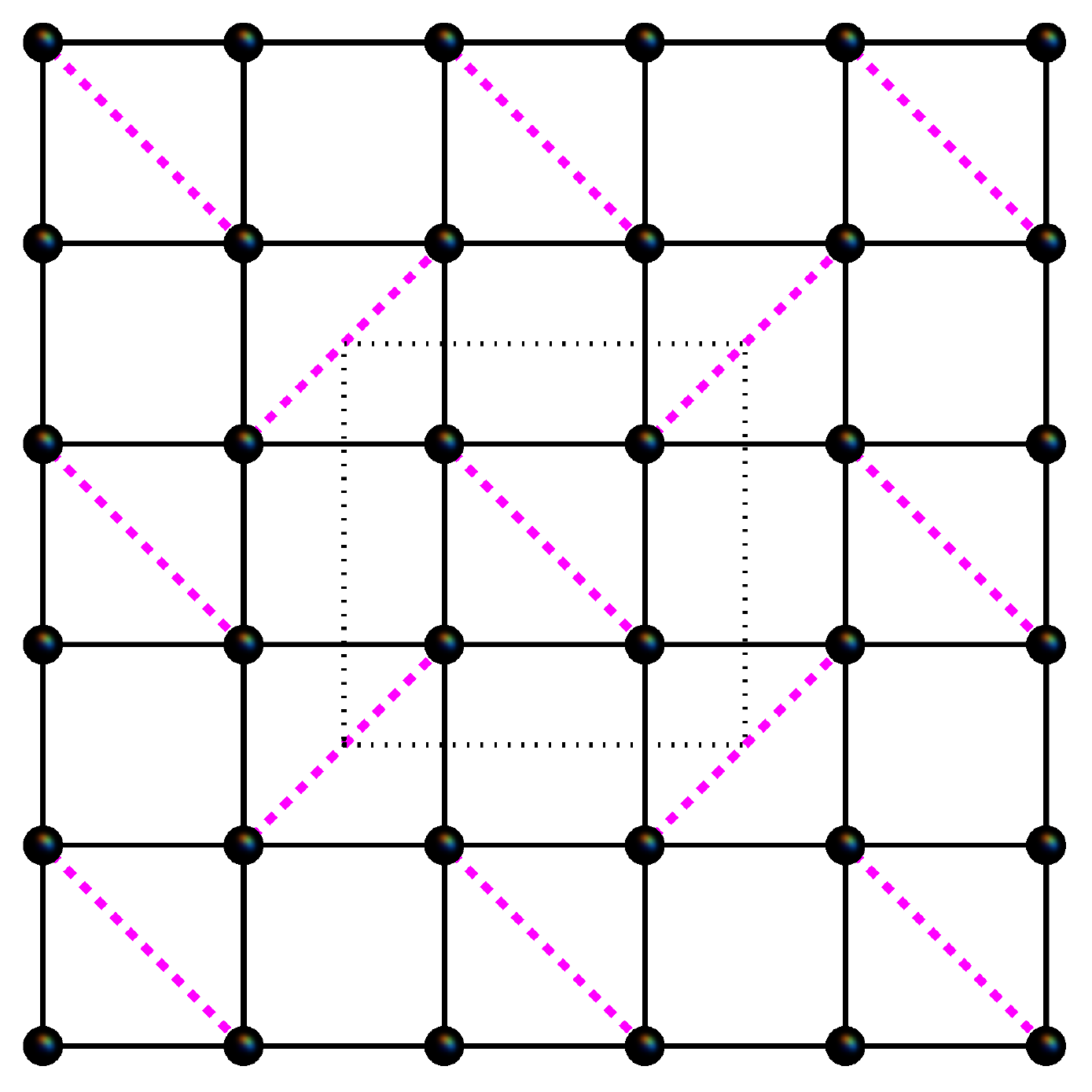}
\caption{(color online) {\bf $Z_4$ electric field configuration at $\nu=2$ on the SSL.} Here, the Gauss law requires $\underset{\bs r ' \in \la \bs r \bs r' \ra}{\prod} e_{\bs r \bs r'} = -1$ on each site. Solid (Black) and dashed (magenta)  links indicate electric field $e_{\bs r \bs r'} = 1$ and -1 respectively. Note that all space-group symmetries are preserved by this pattern.}
\label{fig:SSLEfieldNu2}
\end{figure}

\begin{figure}
\includegraphics[width=0.5\columnwidth]{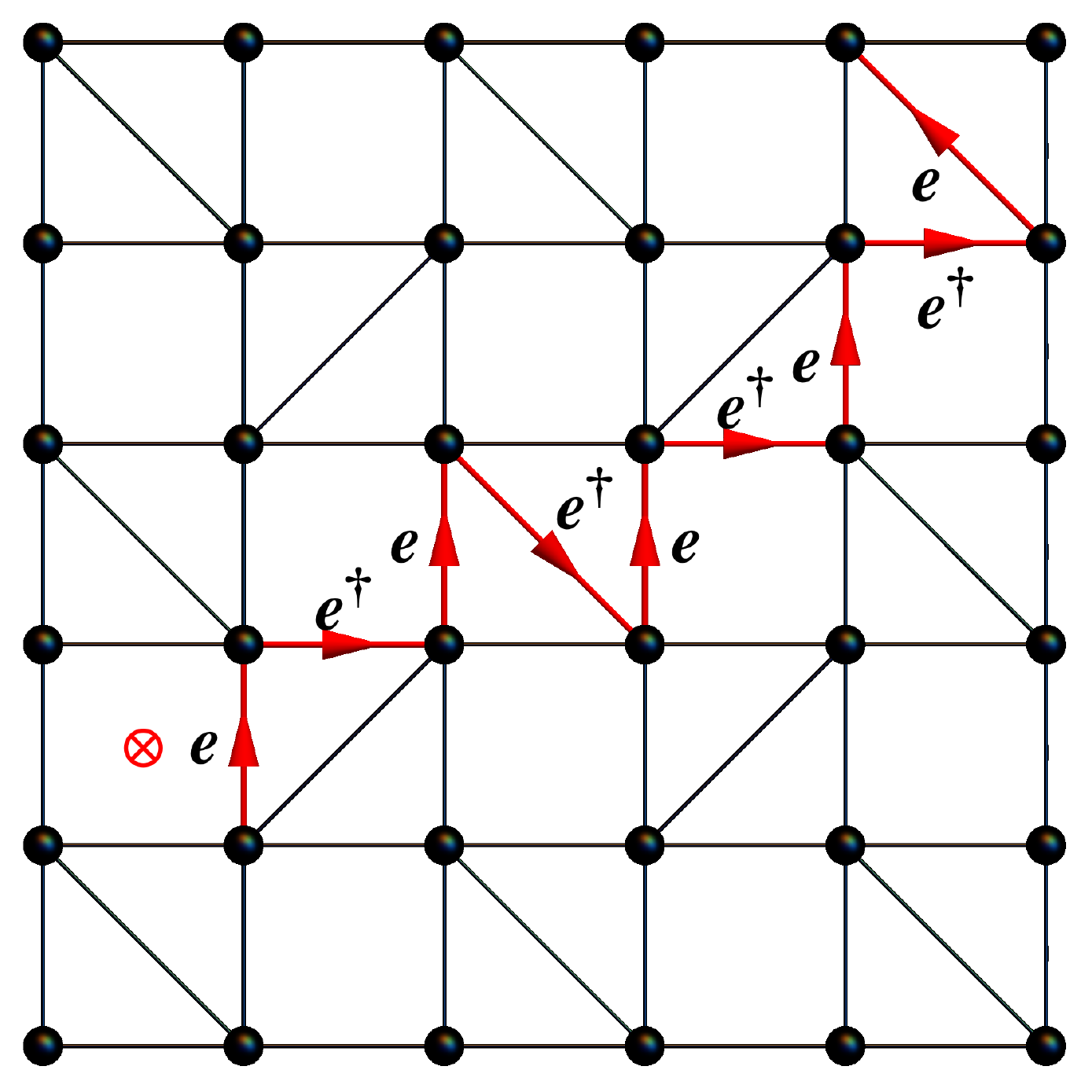}
\caption{(color online) Vison creation operators at a square plaquette $\otimes$: products of electric field operator $e_{\bs r \bs r'}$ and $e_{\bs r \bs r'}^\dagger$ along arrow lines.   }
\label{fig:FluxInsSSL}
\end{figure}

\subsection{Dual $Z_4$ clock model} 
\label{subsec:z4-clock}
As in the Ising case, the dual theory is a convenient language to study the vison, as the nonlocal duality mapping renders the vison creation operator a local object. To that end, we introduce a 
 a new set of $Z_4$ operators $E_{\bsbr}$ and $A_{\bsbr}$ that reside on each site $\bsbr$ of the dual pentagon lattice. These variables have similar Hilbert space structure as in (\ref{eq:linkactionZ4}) and are related to  $a, e$ via
\bea
e_{\bs r \bs r'} &=& \eta_{\bsbr \bsbr'} A^\dagger_{\bsbr} A_{\bsbr'}
\label{eq:z4dualA} \\
\underset{\bs r  \bs r' \in \Box, \vartriangle }{\prod} a_{\bs r \bs r'} &=& E_{\bsbr} ,
\label{eq:z4dualE}
\eea

The Gauss law constraint in the original lattice site maps to the product of $\eta_{\bsbr \bsbr '} $ values for every pentagon: $\underset{\bs r' \in \la \bs r  \bs r' \ra}{\prod} e_{\bs r \bs r'}= \underset{\bsbr \bsbr' \in \pentagon}{\prod} \eta_{\bsbr \bsbr'}$. The dual theory then takes the form of a $Z_4$ clock model on the pentagonal lattice,
\be
\mathcal{H}_{\bar g}= -h \sum_{\bsbr \bsbr'} \eta_{\bsbr \bsbr'} A^\dagger_\bsbr A_{\bsbr '} 
 -K \sum_{\bsbr} E_{\bsbr}  + \text{h.c.} ,
\label{eq:Z4clock}
\ee
where the bond strengths $\eta_{\bsbr\bsbr'}$ satisfy
\be
\underset{\bsbr \bsbr' \in \pentagon}{\prod} \eta_{\bsbr \bsbr'} = i^\nu
\label{eq:frustratedbondZ4clock}
\ee
at boson filling $\nu$. The non-trivial product of bond variables around a plaquette for $\nu=1$ indicate that the clock model is {\it frustrated}\footnote{Note that we may verify that the flux pattern preserves symmetry for $\nu=2$.}. Note that we can readily construct a bond configuration satisfying  (\ref{eq:frustratedbondZ4clock}) by examining the electric field configurations in Figs.~\ref{fig:SSLbase},~\ref{fig:SSLEfieldNu2}, and associating the value of $\eta_{\bsbr\bsbr'}$ on a bond of a dual lattice with the value of the electric field on the direct lattice bond bisected by $\bsbr\bsbr'$. Once this assignment is made, the couplings $\eta_{\bsbr\bsbr'}$ are held fixed, i.e. they are not dynamical objects.

In the dual theory, the vison creation operator $\hat{F}_\otimes$ defined by (\ref{eq:visonstringoperatorZ4}) is represented via
\be
\hat{F}_\otimes = \Big( \underset{\bsbr \bsbr' \rightarrow \infty }{\prod}   \eta_{\bsbr \bsbr '}^*  \Big)A_1.
\ee
Although this includes a non-local string product of the bond strengths $\eta_{\bsbr\bsbr'}$, as we have already noted, these are fixed and non-dynamical. Thus, as promised, in the dual theory $\hat{F}_\otimes$ is a local operator; we now study its symmetry properties.
\begin{figure}
\includegraphics[width=0.6\columnwidth]{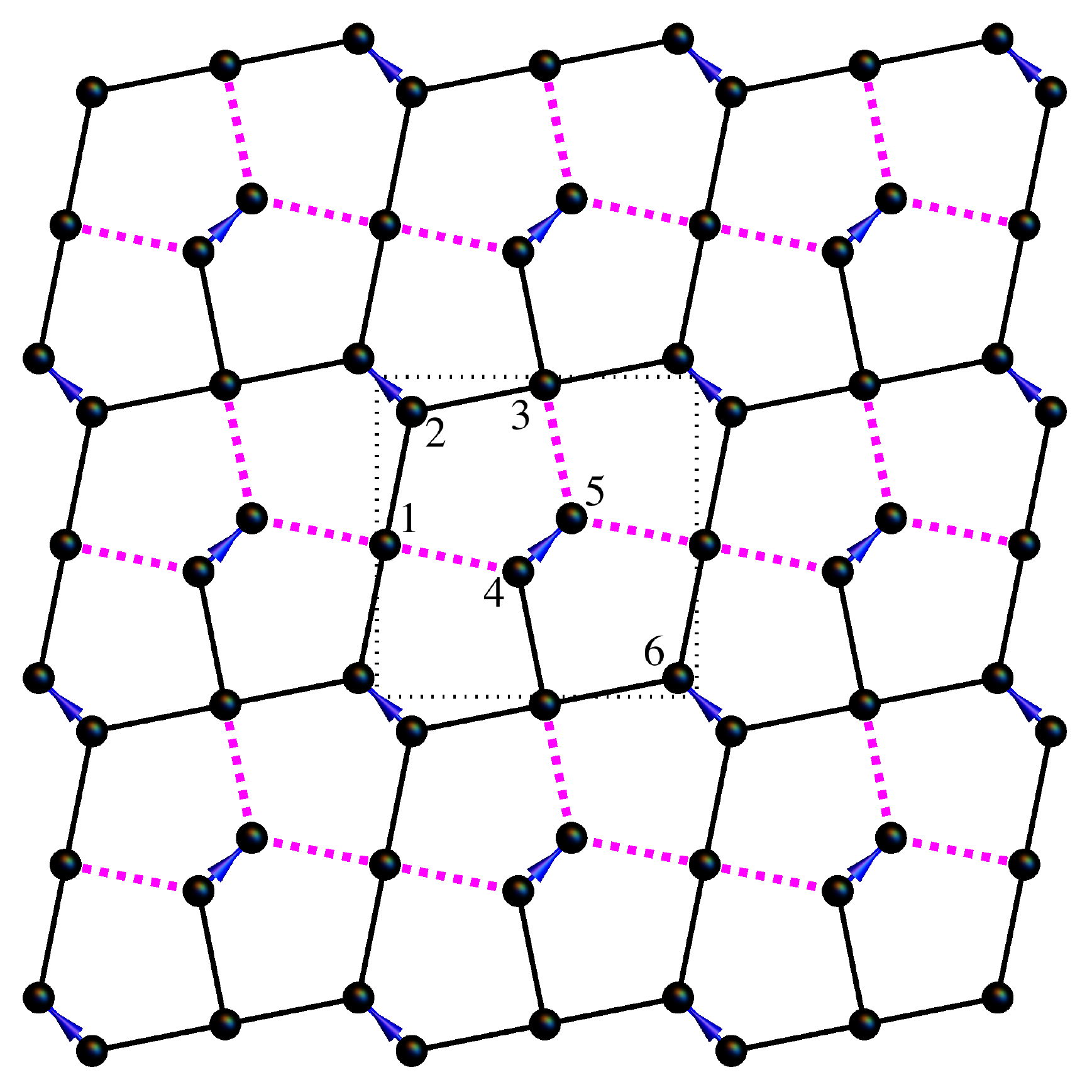} 
\caption{(color online) {\bf Configuration of bond signs $\eta_{\bsbr \bsbr'}$ on the dual pentagonal lattice for $\nu=1$}. At this filling, 
 $\underset{\bsbr \bsbr' \in \pentagon}{\prod} \eta_{\bsbr \bsbr'} = i $ (link products are taken in the anti-clockwise sense).  Solid (Black) and dashed (magenta) links indicate electric field $\eta_{\bsbr \bsbr'} = 1$ and -1 respectively. Blue arrows on each diagonal link represents $\eta_{\bsbr \bsbr'} = i$. The dotted square shows a single six-site unit cell. Note that the pattern breaks point group symmetry but not translations, as expected.}
\label{fig:SSLdualnu1}
\end{figure}

\subsection{Vison symmetry analysis} 
In order to study the fractionalization of space group symmetries, we now consider the transformation of the $A_{\bsbr}$ under  lattice symmetries.

The symmetries of the group $p4g$  (shared by both the SSL and the dual pentagonal lattice) are generated by the following operations:
translations along orthogonal lattice primitive vectors ($\bs a_1 \equiv (2,0)$ and $\bs a_2 \equiv (0,2)$):
\bea
T_{\bs a_1} &:& (x,y) \mapsto (x,y) + \bs a_1 
 \nn\\
T_{\bs a_2 } &:& (x,y) \mapsto (x,y) + \bs a_2, 
\label{eq:p4gtrans}\eea
mirror reflections along axes oriented at $\pi/4$ with respect to the lattice vectors:
\bea
\sigma_{xy} &:& (x,y) \mapsto (y,x)
\nn\\
\sigma_{x\bar{y}} &:& (x,y) \mapsto (-y , -x), 
\label{eq:p4gmirror1}
\eea
and glide reflections about  axes parallel to the lattice vectors:
\bea
G_{x} &:& (x,y) \mapsto (x,-y) + \frac{1}{2}\left(\bs a_1-\bs a_2\right) 
\nn\\
G_{y} &:& (x,y) \mapsto (-x,y) -\frac{1}{2}\left(\bs a_1-\bs a_2\right),
\label{eq:p4gmirror} 
\eea
(See Fig.~\ref{fig:SSLbase} for the lattice structure and lattice vectors.) Note that we have chosen a center of symmetry that renders $\sigma_{xy},\sigma_{x\bar{y}}$ very simple and underscores that they do not involve any translations, at the cost of making the glide operation slightly more involved. The crucial point is that the associated translations are not  projections of lattice vector onto the glide planes, a fact that guarantees that the glide can not be removed by a suitable change of origin~\cite{KonigMerminPNAS}.

These transformation properties map the values of field operators at different lattice sites into each other. The transformation properties of the $A_{\bsbr}$ depends crucially on the set of $\eta_{\bsbr\bsbr'}$ and hence implicitly on the original boson filling. For $\nu =2$, the $\eta_{\bsbr \bsbr'}$ configuration does not break any of space group symmetries, and therefore it is a straightforward exercise to show that $A_{\bsbr}$ transforms trivially under lattice symmetries.  In contrast, for $\nu=1$  the assignment of $\eta_{\bsbr\bsbr'}$ satisfying  $\underset{\bsbr \bsbr' \in \pentagon}{\prod} \eta_{\bsbr \bsbr'} = i $ necessarily breaks point-group symmetries and therefore $A_{\bsbr }$ transforms projectively. In order to determine the transformation laws of the $A_{\bsbr }$  under symmetry, it suffices to consider how the transformations (\ref{eq:p4gtrans}-\ref{eq:p4gmirror}) act on $A_{\bsbr }$ while keeping the combination $\sum_{\bsbr\bsbr'} \eta_{\bsbr\bsbr'} A_{\bsbr}^\dagger A_{\bsbr'} + \text{h.c.}$ invariant. {This amounts to constructing the {\it projective symmetry group} in the standard terminology of the parton construction of fractionalized phases\cite{WenPSG}.} 

First, note that it is straightforward to see that the $A_{\bsbr}$ transform trivially under translations $T_{\bs a_1}$ and $T_{\bs a_2}$ since $\eta_{\bsbr \bsbr'}$ phases do not enlarge the unit-cell. We may therefore consider only the point-group symmetries. It is useful to introduce some notation: let us label the unit cells by integers $(x,y)$ such that $\bsbr_{(m,n)} = m \bs a_1 + n \bs a_2$ and label the six dual lattice sites within a single unit cell as shown in Fig.~.
By examining how the action of the four symmetries ($\sigma_{xy}$, $\sigma_{x\bar{y}}$, $G_x$, $G_y$) relates these six sublattice indices while simultaneously transforming the unit cell coordinates we arrive at 
Table~\ref{tab:visonSymmetryZ4}. 
 As an example of how to construct the entries in Table~\ref{tab:visonSymmetryZ4}, let us consider the reflection $\sigma_{xy}$. 
 Following (\ref{eq:fieldtransformations}), we see that under this operation, we have
 \be
 \sigma_{xy}: e_{\bs r \bs r'} \rightarrow e_{\sigma_{xy}(\bs r) \sigma_{xy}(\bs r')}
 \ee
and we must choose the action of $\sigma_{xy}$ on $A_{\bsbr}$ to be consistent with this, given the expression (\ref{eq:z4dualA}) relating $e_{\bs r \bs r'}$ and $A_{\bsbr}$. First, observe that we must have $A_{\bsbr}\rightarrow \lambda_{\bsbr}A^\dagger_{\sigma_{xy}(\bsbr)}$ in order for the transformed $e_{\bs r \bs r'}$ to be proportional to $e_{\sigma_{xy}(\bs r) \sigma_{xy}(\bs r')}$; the phases are then fixed by requiring $\mathcal{H}_{\bar g}$. From Fig.~~\ref{fig:SSLdualnu1}, we see that under this symmetry, the sublattices transform via $1\leftrightarrow 3$, $2\leftrightarrow 6$, $4\rightarrow 4$ and $5\rightarrow 5$. 
Furthermore, note that owing to the phase difference  $\eta_{14} = - \eta_{34} =1$, we must require that $\sigma_{xy}$ induce a sign change only on sublattice 4 so that $\eta_{\bsbr\bsbr'} A_{\bsbr}^\dagger A_{\bsbr'} $ remains invariant. Proceeding in this fashion, we may construct the other entries in Table~\ref{tab:visonSymmetryZ4}.  
 \begin{table*}
\centering
\begin{tabular}{|  l  || l  | l | l | l  |} 
\hline 
 \phantom{we need a } &~~~ $\sigma_{xy} $ & $ ~~~~~ \sigma_{x\bar{y}} $&  $~~~~~~ G_x $ & $~~~~~~ G_y$  \\
 \hline \hline
$~~A_{(m,n) 1}~~ $ & $ \phantom{-}A_{(n, m-1) 3 }^\dagger $  & $ \phantom{-} (-1)^{m+n } A_{(-n , -m) 3}^\dagger $
& $  i^{2n+1} A_{(m , -n-1) 3}^\dagger $ & $ i^{2m+1} A_{(-m,n) 3}^\dagger  $ \\ 
\hline
$~~A_{(m,n) 2}~~ $ & $ \phantom{-} A_{(n,m) 6 }^\dagger $  & $ \phantom{-}(-1)^{m+n } A_{(-n , -m) 2}^\dagger $
& $  i^{2n+3} A_{(m , -n-1) 5}^\dagger $ & $ i^{2m+1} A_{(-m,n+1) 4}^\dagger  $ \\ 
\hline 
$~~A_{(m,n) 3}~~ $ & $ \phantom{-} A_{(n+1, m) 1 }^\dagger $  & $ \phantom{-} (-1)^{m+n} A_{(-n , -m) 1}^\dagger $
& $  i^{2n+1} A_{(m+1 , -n-1) 1}^\dagger $ & $ i^{2m+3} A_{(-m,n+1) 1}^\dagger  $ \\ 
\hline 
$~~A_{(m,n) 4}~~ $ & $ - A_{(n, m) 4 }^\dagger $  & $ \phantom{-} (-1)^{m+n } A_{(-n , -m) 5}^\dagger $
& $  i^{2n+3} A_{(m , -n) 6}^\dagger $ & $ i^{2m+3} A_{(-m,n) 2}^\dagger  $ \\ 
\hline 
$~~A_{(m,n) 5}~~ $ & $ \phantom{-} A_{(n, m) 5 }^\dagger $  & $\phantom{-}  (-1)^{m+n } A_{(-n , -m) 4}^\dagger $
& $  i^{2n+3} A_{(m+1 , -n-1) 2}^\dagger $ & $ i^{2m+1} A_{(-m-1,n+1) 6}^\dagger  $ \\ 
\hline 
$~~A_{(m,n) 6}~~ $ & $ \phantom{-} A_{(n, m) 2 }^\dagger $  & $ - (-1)^{m+n} A_{(-n , -m) 6}^\dagger $
& $  i^{2n+1} A_{(m+1 , -n) 4}^\dagger $ & $ i^{2m+1} A_{(-m-1,n) 5}^\dagger  $ \\ 
\hline
\end{tabular}
\caption{{\bf Vison symmetries}. We list the transformation of  the single-vison operator $A_{\bsbr }$ under four lattice symmetries, for a given phase configuration $\eta_{\bsbr \bsbr'}$ that satisfies $\underset{\bsbr \bsbr' \in \pentagon}{\prod} \eta_{\bsbr \bsbr'} = i $ (see Fig.~\ref{fig:SSLdualnu1}) }
\label{tab:visonSymmetryZ4}
\end{table*}
\begin{figure}
\includegraphics[width=0.6\columnwidth]{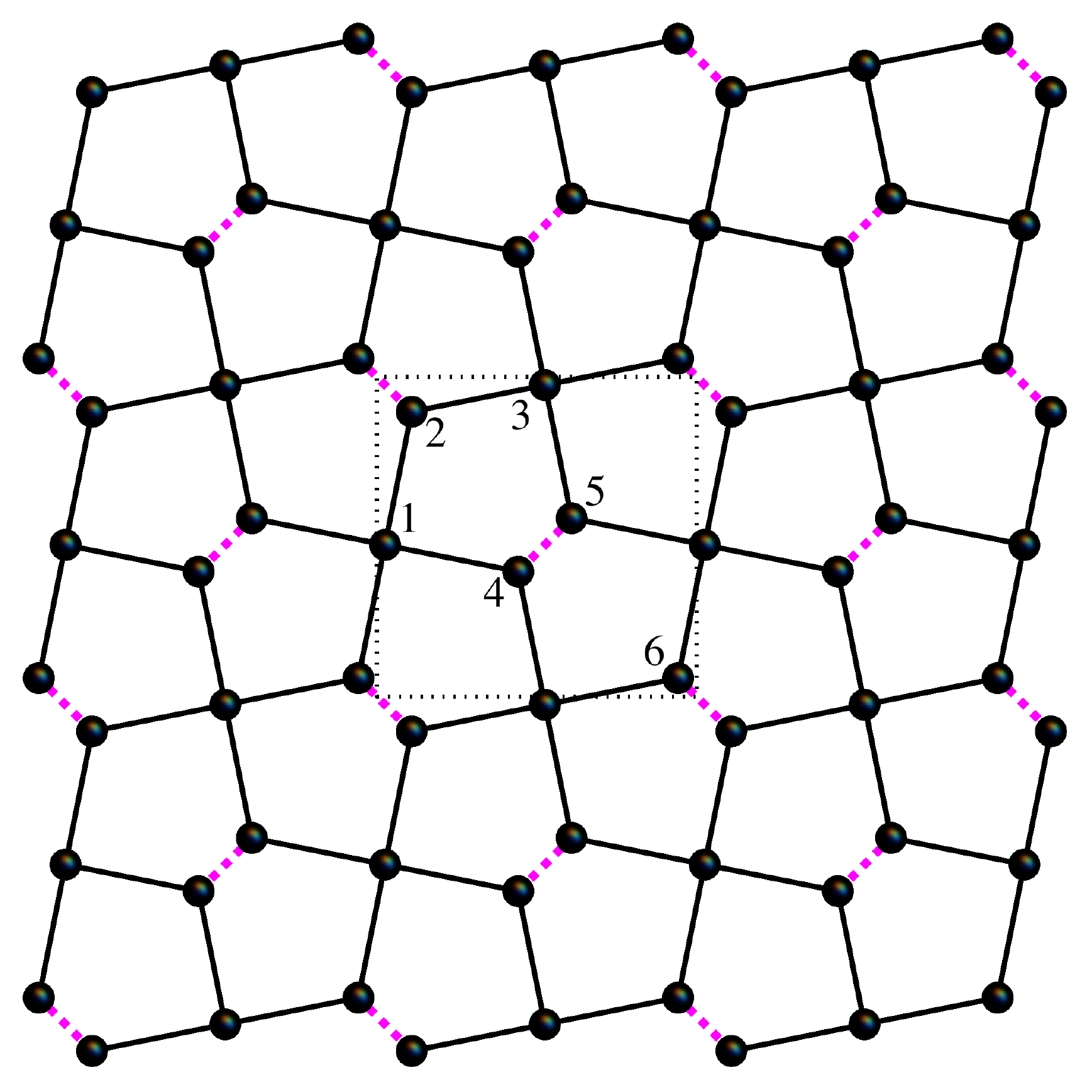}
\caption{(color online) {\bf Configuration of bond signs $\eta_{\bsbr \bsbr'}$ on the dual pentagonal lattice for $\nu=2$}. At this filling,
 $\underset{\bsbr \bsbr' \in \pentagon}{\prod} \eta_{\bsbr \bsbr'} = -1 $. 
 Solid (Black) and dashed (magenta) links indicate electric field $\eta_{\bsbr \bsbr'} = 1$ and -1 respectively. Note that the pattern breaks no symmetries.}
\label{fig:SSLdualnu2}
\end{figure}

Table~\ref{tab:visonSymmetryZ4} allows us to  compute relations between different symmetries when acting on single-vison states. Operationally, we may obtain these relations by constructing the state $\ket{v_\bsbr} \equiv A^\dagger_{\bsbr} \ket{0}$ and acting upon it with the different symmetry operators in turn.  
First, we find that a subset of the space group symmetries satisfy a `trivial' algebra, in that they do not exhibit any difference when acting on single visons compared to their multiplication table computed within the space group (without reference to the vison states):
\begin{subequations}
\bea
T^v_{\bs a_1} T^v_{\bs a_2}  &=& T^v_{\bs a_2} T^v_{\bs a_1}
\label{eq:normalSGprodZ4_a} \\
(\sigma_{xy}^v)^2  &=& 1
\label{eq:normalSGprodZ4_b} \\
(\sigma_{x\bar{y}}^v)^2 &=& 1
\label{eq:normalSGprodZ4_c} \\
G_y^v  \sigma_{xy}^v  &=& \sigma_{xy}^v G_x^v
\label{eq:normalSGprodZ4_d} \\
T_x^v (G_y^v)^{-1} &=& \sigma_{x\bar{y}}^v G_x \sigma_{x\bar{y}} ^v
\label{eq:normalSGprodZ4_e} 
\eea
\end{subequations}
where the `$v$' denotes the fact that we are considering the action on single-vison states.
In contrast, the remaining set of relations between the space group symmetry generators includes a projective phase factor of (-1) relative to their expected forms:
\begin{subequations}
\bea
(G_x^v)^2 &=& -T^v_{\bs a_1} 
\label{eq:anomalousSGprodZ4_a}\\ 
(G_y^v)^2 &=& -T^v_{\bs a_2} 
\label{eq:anomalousSGprodZ4_b}\\ 
\sigma^v_{x\bar{y}} G_x^v &=& - G_x^v \sigma_{xy}^v 
\label{eq:anomalousSGprodZ4_c}\\ 
\sigma^v_{x\bar{y}} \sigma_{xy}^v &=& - \sigma_{xy}^v \sigma_{x\bar{y}}^v 
\label{eq:anomalousSGprodZ4_d}\\ 
G_x^v T_{\bs a_2}^v &=& - (T^v_{\bs a_2})^{-1} G_x^v. 
\label{eq:anomalousSGprodZ4_e}
\eea
\end{subequations}
The non-trivial (-1) phase factor that appears in the above algebraic relations cannot be gauged away by redefinitions of the symmetry operations. This is once again an indication that the visons fractionalize symmetry: in this case, the fractionalized symmetry corresponds to the glide planes (and the remaining nontrivial relations should be viewed as consequences of this.) We may readily confirm that for $\nu=2$, such a phase factor is absent: there is no point group symmetry fractionalization. Indeed, our arguments may be straightforwardly extended to all fillings, and we find  (perhaps unsurprisingly!) that the relevant phase factor is $(-1)^{\nu\,(\text{mod }\mathcal{S})}$, so that glide quantum number fractionalization only occurs for fillings that are not a multiple of the non-symmorphic rank.

\subsection{Condensing $Z_4$ fluxes: confined phases and $Z_2$ gauge theories }
\label{subsec:vison-condensation}
In the previous section, we have studied the symmetries of single vison excitations in the deconfined phase of  $\mathcal{H}_g$ that emerges in the limit $K \gg h$. We have demonstrated that at odd integer filling the visons fractionalize point-group symmetries while preserving translational symmetries. We now  focus on the  case of $\nu=1$, and analyze the proximate phases that can be accessed from our $Z_4$ theory by condensing visons. In a $Z_4$ theory, we may choose to condense either one, two, or three visons; each of these leads to distinct possibilities. Note that we do {\it not} construct the specific microscopic Hamiltonians needed to drive the system into  these vison-condensed phases; we simply use the preceding symmetry analysis to draw universal conclusions about the symmetry and topological properties of the vison condensates. 

A crucial fact is that  condensing  particles in the deconfined phase of a gauge theory confines all particles that have non-trivial mutual statistics with the condensate, but leaves particles with trivial mutual statistics as deconfined excitations. The charges (denoted $e)$ and fluxes (denoted $m$) in the $Z_4$ gauge theory take values $q_e, q_m\in \{0,1,2,3\}$, with the mutual statistics phase factor for taking an $e$-particle around an $m$-particle (or vice-versa) given by $e^{2\pi i q_e q_m/4}$. We will now study the phases obtained by the different possibilities for vison condensation.

\noindent{\bf Single Vison Condensation:} Imagine we exit the deconfined phase of $\mathcal{H}_g$ by condensing a single vison. The resulting phase will have  $\langle{A}_{(x,y)\mu}\rangle \neq 0$ for some $\mu \in \{1,2,\ldots,6\}$.  Since the single-vison state corresponding to $A$ has $q_m =1$, it follows that in a single-vison condensate, all fluxes are identified with the vacuum (since it is a condensate of fluxes), and all the $e$-particles are confined, since they all have nontrivial statistics with a $q_e=1$ object. Thus, condensing the vison results in confinement of the $Z_4$ gauge field, and the symmetry relations (\ref{eq:anomalousSGprodZ4_a}-\ref{eq:anomalousSGprodZ4_e}) reveal that the system breaks point-group symmetries, owing to the nonzero value of $\langle{A}_{(x,y)}^\mu\rangle$. We note that condensing visons provides a convenient unified formalism for examining broken-symmetry states on the SSL.

\noindent{\bf Vison Pair Condensation: } A more interesting situation arises if  energetics favor the condensation of {\it paired} visons over the single-vison condensate.
We may understand the nature of the resulting phase as follows. As $A^2$ carries two units of magnetic flux ($q_m=2$), creating a two-vison condensate identifies $q_m=2$ with $q_m=0$ and hence $q_m=1$ with $q_m=3$: in other words, the fluxes now take values in the group $Z_2$. Now, we see that condensing a $q_m=2$ object must confine the $q_e =1$ and the $q_e=3$ charges, as they have nontrivial mutual statistics with it; however, the $q_e =2$ charge remains deconfined.  Thus, the charges also take values in $Z_2$, and we are left with a $Z_2$ gauge theory. As the $q_e=2$ charge must be equivalent to a two-parton bound state, we conclude that it also carries $1/2$ charge of the global $U(1)$ symmetry. If we can construct a vison-paired state without breaking symmetry, then we will arrive at a simpler fractionalized symmetry-preserving phase of $\nu=1$ bosons on the SSL; this would be a phase with deconfined quasiparticles with $1/2$-charge under the global $U(1)$ symmetry, and emergent $Z_2$ gauge flux. 

In order to construct such a state, it suffices to consider nearest-neighbor vison pair operators to identify appropriate combination(s) of visons that preserve all symmetries. Intuitively, we wish to identify a vison pair such that acting with any  symmetry operator leads to a {\it squaring} of phase factors in (\ref{eq:anomalousSGprodZ4_a}-\ref{eq:anomalousSGprodZ4_e}) so that the  (-1) factors  are absent: in other words, we wish to identify a  two-vison condensate that transforms trivially under symmetries. 

To this end, we look for a vison pair operator 
$\sum_{\langle r r' \rangle } (c_{r r'} A_r A_{r'} +\text{h.c.}) $ that is invariant under all the symmetries. 
Using the vison transformations listed in Table \ref{tab:visonSymmetryZ4} we identify the combination 
$A_1A_2 + A_3A_6 + A_1 A_4 - A_3 A_4 + A_2 A_3 + A_1A_6 + A_3 A_5 + A_1 A_5$ that satisfies this requirement
(we have suppressed spatial indices except for those that denote the sublattice, for simplicity.) 
Thus, it is indeed possible to condense a vison pair without breaking any of the point group symmetries.

In summary:  upon condensing the vison pair state identified above we arrive at a $Z_2$ gauge theory, where the single vison continues to transform projectively under the point-group symmetries, and the single chargon carries $1/2$ unit of the global $U(1)$ charge. In Appendix~\ref{app:dorbitalglidefrac} we show that a $Z_2$ gauge theory with the same symmetry structure arises in a different model with the same $p4g$ and $U(1)$ symmetries and filling of $\nu=1$, but with two sites per unit cell, underlining the fact that the structure is universal to the symmetry group rather than any particular tight-binding lattice model.

\noindent{\bf Vison Triplet condensation:} Condensing a three-vison bound state once again leads to a confined phase with broken symmetry, by an argument analogous to the one-vison case. First, observe that all the $e$ particles have nontrivial mutual statistics and are hence confined. Now, we have identified $q_m =3$ and the vacuum, $q_m \equiv 0$; but since we began with a $Z_4$ gauge theory, $q_m=4$ is {\it also} the vacuum. We can check that this guarantees that the particles with $q_m =1,2$ are also identified with the vacuum\footnote{To see this, first observe that combining particles with $q_m=1$ and $q_m=3$ in a $Z_4$ gauge theory should be equivalent to the vacuum, but since the $q_m=3$ particle is condensed and is hence equivalent to the vacuum, it follows that the $q_m=1$ particle must also be condensed. A similar argument based on the fact that combining $q_m=3$ and $q_m=2$ should give $q_m=1$ reveals that $q_m=2$ is also condensed.} so there are no magnetic flux excitations. Finally, we observe that a triplet of fluxes will also carry a (-1) factor for the symmetry operations (\ref{eq:anomalousSGprodZ4_a}-\ref{eq:anomalousSGprodZ4_e}), and hence this condensate breaks symmetry.

\section{\label{sec:detection}Detecting Glide Fractionalization}
\subsection{\label{subsec:numerics}Numerical Signatures}
We now turn to a discussion of how to detect glide symmetry fractionalization in numerics. As in the rest of the paper, we will focus on the case of $Z_2$ topological order; note that while our initial construction for a topologically ordered phase for the $s$-orbital Shastry-Sutherland model invoked a $Z_4$ gauge structure with three distinct vison excitations, by condensing paired visons we were able to access a theory with $Z_2$ topological order with no broken symmetry, where the single remaining vison excitation continues to transform nontrivially under glides. We assume that the topological order is of the `toric code' type invoked in the preceding discussion and do not consider the alternative `doubled semion' theory here. 

We follow a line of reasoning developed by Zaletel, Lu and Vishwanath~\cite{ZaletelPSGMeasurement} (ZLV), and use similar notation where possible (Essentially identical results were obtained, albeit from a somewhat different perspective, by Qi and Fu~\cite{QiFu}.) Recall that a $Z_2$ topologically ordered phase has a four-fold degeneracy on an infinite cylinder. In line with ZLV, we assume the existence of a numerical procedure that can generate the ground-state manifold in the basis of `minimally entangled states' (MESs). The MES basis is the unique basis for the ground-state manifold in which the unitary operation of `threading anyonic flux $a$',  denoted $\mathcal{F}^a_x$, is realized as a permutation of basis states. Formally, $\mathcal{F}^a_x$  relates ground states to ground states by creating a pair of anyons $(a, \bar{a})$ from vacuum and dragging in opposite directions $\pm \hat{x}$ to infinity; each MES thus has a unique topological flux threading the cylinder. Let us denote the four anyon types of the toric code as follows: `$1$' (the vacuum and any local excitations, that must carry integer global $U(1)$ charge); the bosonic chargon `$b$', carrying $1/2$ charge; the vison `$v$', carrying fractional glide quantum number; and  the fermion`$f$', the composite object $f=bv$. The vison, as discussed, acts as a $\pi$ flux for the chargon. (In the $Z_2$ gauge theory language used in the preceding sections, $b$ is the electric charge $e$, $v$ is the magnetic flux $m$ and $f$ is the dyonic bound state of the two.) Strictly speaking, anyon types label not the MES themselves,  but only label {\it differences} between MESs~\cite{ZaletelPSGMeasurement}. However, this subtlety is not important for our purposes and we ignore it here.

In the example we study here, the non-trivial quantum numbers are associated with the single-vison sector; we will focus on this case. As we will show, we may reduce the set of nontrivial relations into just three operations that square to $-1$ when acting on single-vison states: (i) site-centered inversion; (ii) a generalized `mirror' symmetry on odd cylinders; and (iii) bond-centered inversion. As noted by ZLV, it is particularly easy to identify symmetry quantum numbers  of an operator $\mathcal{O}$ that (like the three listed above) exchanges the two ends of a finite cylinder. Consider creating a pair of vison excitations\footnote{Since we focus on $Z_2$ phases, where each particle is its own antiparticle, we omit the bar on antiparticles henceforth.} $(v, {v})$  from the vacuum that are related by any such edge-exchanging operator $\mathcal{O}$; we can view the final state as a different MES, corresponding to the insertion of a flux of $v$ through the original MES. If the eigenvalue of $\mathcal{O}$ changes sign during this process, then the pair of anyons is {\it odd} under the operation $\mathcal{O}$. Since both visons together transform with eigenvalue $-1$, each individual vison carries the `fractionalized' $\mathcal{O}$ quantum number of $\sqrt{-1}$. 
 
We now briefly summarize the approach developed by ZLV to measure this symmetry fractionalization. Consider a large finite cylinder with an operator $\mathcal{O}$ that exchanges its edges.  Take a state $\ket{\Lambda, a}$ that is symmetric under $\mathcal{O}$, where $\Lambda$ denotes geometric details such as the dimensions, edge structure, etc., and $a \in \{1, b,v,f\}$ labels the topological sector of the cylinder, and the bulk is assumed to be in some definite topologically ordered phase. Then, the global quantum number of $\mathcal{O}$ is defined by
\be
\mathcal{O}\ket{\Lambda, a} = Q_{\mathcal{O}}(\Lambda, a)\ket{\Lambda, a}.\label{eq:globalOquantdef}
\ee
Although the global quantum number $Q_{\mathcal{O}}$ is insensitive  deformations at the edge or the bulk that preserve the symmetry and the bulk topological order\cite{ZaletelPSGMeasurement}, it can depend on the topological sector: in order to toggle between different topological sectors, we need to create an anyon pair $(v,v)$ from the vacuum and pull them to the edge, in effect acting with $\mathcal{F}_x^v$ on our original state: in other words, we have
\be
\ket{\Lambda, a\cdot v} = \mathcal{F}^v_x \ket{\Lambda, a}.
\ee
Therefore, an invariant, $\Lambda$-independent definition of the quantum number is provided by the {\it relative} quantum number between topological sectors related by a flux insertion:
\be
Q^{(v)}_{\mathcal{O}} \equiv \frac{Q_{\mathcal{O}}(\Lambda, av)}{Q_{\mathcal{O}}(\Lambda, a)}
\ee
where the numerator and denominator are each defined by (\ref{eq:globalOquantdef}) with the appropriate choice of MES. Thus, for any operation $\mathcal{O}$ that squares to $-1$ when acting on a single-anyon (in our case, single-vison) state and exchanges the edges, we may detect this symmetry fractionalization in a numerical experiment by computing $Q^{(v)}_{\mathcal{O}}$ according to the preceding discussion and verifying that it yields $-1$.

\subsection{\label{subsec:opid}Identifying Fractionalized Symmetry Operators}
We now turn to identifying the three possible such edge-exchanging operators that are fractionalized at odd $\nu$ in the $p4g$ space group. As a first step, we reduce the set of nontrivial relations. For this purpose, it is convenient to work with a slightly different set of glides $\tilde{G}_{x,y}$, related to the original glides (\ref{eq:p4gmirror}) via $\tilde{G}_x \equiv G_x T_{\bs a_2}^{-1}$, $\tilde{G}_y \equiv G_y T_{\bs a_1}^{-1}$. Note that this allows us to express all other symmetry generators solely in terms of $\tilde{G}_x$ and $\sigma_{xy}$: 
\bea
T_{{\bs a}_1} &=& \tilde{G}_x^2, \nonumber\\
\tilde{G}_y &=& \sigma_{xy}\tilde{G}_x \sigma_{xy},\nonumber\\
T_{\bs a_2} &=& \tilde{G}_y^2,\nonumber\\
\sigma_{x\bar{y}} &=& \tilde{G}_x^{-1} \sigma_{xy} \tilde{G}_x\label{eq:operatorredef}. 
\eea
 Working in terms of these redefined symmetry operators and exploiting the `gauge freedom'~\cite{PhysRevB.87.104406} in defining the  action of symmetry on single anyons (details are provided in Appendix~\ref{app:gaugefixing}) we arrive at a `gauge-fixed' set of normal relations
\begin{subequations}
\bea
(\tilde{G}^v_x)^2 &=&  T^v_{\bs a_1},\label{eq:tildesGxGxTxmod}\\
(\tilde{G}^v_y)^2 &=&  T^v_{\bs a_2},\label{eq:tildesGyGyTymod}\\
\tilde{G}_y^v &=& \sigma_{xy}^v\tilde{G}_x^v \sigma_{xy}^v,\label{eq:tildesGydefmod}\\\
\sigma_{x\bar{y}}^v &=& (\tilde{G}_x^v)^{-1} \sigma_{xy}^v \tilde{G}^v_x,\label{eq:sigmaxbarydefmod}\\
T^v_{\bs a_1}T^v_{\bs a_2}&=& T^v_{\bs a_2}T^v_{\bs a_1} ,\\
(\sigma^v_{xy})^2 &=& (\sigma^v_{x\bar{y}})^2 = 1\label{eq:sigmasquaredmod},
\eea 
\end{subequations}
and  nontrivial ones,
\begin{subequations}
\bea
(\sigma_{xy}^v\sigma^v_{x\bar{y}})^2 &=& -1\label{eq:redanom1},\\
\tilde{G}^v_x T_{\bs a_2}^v &=& - (T_{\bs a_2}^v)^{-1} \tilde{G}^v_x\label{eq:redanom2},\\
(T_{\bs a_1}^v)^{-1} (\tilde{G}_y^v)^{-1} &=& -\sigma^v_{x\bar{y}} \tilde{G}_x^v \sigma^v_{x\bar{y}}\label{eq:redanom3}.
\eea
\end{subequations}
Any remaining (-1) signs in these expressions cannot be removed by redefining the gauge, and are hence `physical'. We now turn to identifying signatures of these three nontrivial symmetry relations.

From the set of relations (\ref{eq:redanom1}-\ref{eq:redanom3}), we may identify the three edge-exchanging operators whose quantum number fractionalization may be measured by suitable manipulations on an MES basis, as outlined above. First, observe that the operation $\sigma_{xy}\sigma_{x\bar{y}}$  takes $(x,y)$ to $(-x, -y)$, acting as site-centered inversion operation ($\pi$ rotation in two dimensions) that is clearly edge-exchanging. Thus, (\ref{eq:redanom1})  states that site-centered inversion squares to $(-1)$ acting on single visons.

We next examine the relation (\ref{eq:redanom2}). Consider the operation $\mathcal{M}_m^v \equiv  (\tilde{G}_x^v)^m T_{\bs a_2}^v$, where $m$ is an integer. Using (\ref{eq:redanom2}), 
\be
(\mathcal{M}_m^v)^2 &=& (\tilde{G}_x^v)^m T_{\bs a_2}^v(\tilde{G}_x^v)^m T_{\bs a_2}^v \nonumber\\
				 &=& (\tilde{G}_x^v)^m T_{\bs a_2}^v \tilde{G}_x^v T_{\bs a_2}^v(T_{\bs a_2}^v)^{-1} \tilde{G}_x^v(T_{\bs a_2}^v)^{-1}(\tilde{G}_x^v)^{m-2}T_{\bs a_2}^v\nonumber\\
				 &=& (\tilde{G}_x^v)^{m+2} T_{\bs a_2}(\tilde{G}_x^v)^{m-2}T_{\bs a_2}^v,\label{eq:manipulation}
\ee 
which shifts the powers of the $\tilde{G}_x^v$s by $2$. For even $m$, we use (\ref{eq:manipulation}) $m/2$ times, and find $(\mathcal{M}_m^v)^2 = (\tilde{G}_x^v)^{2m}$; for odd $m$, we use (\ref{eq:manipulation}) $(m-1)/2$ times, and obtain
\be
(\mathcal{M}_m^v)^2 &=& (\tilde{G}_x^v)^{2m-1} T_{\bs a_2}^v\tilde{G}_x^vT_{\bs a_2}^v = - (\tilde{G}_x^v)^{2m} \ee
by using (\ref{eq:redanom2}) again. Thus, we have, using (\ref{eq:tildesGxGxTxmod}), the following equivalent form of (\ref{eq:redanom2}):
\be
(\mathcal{M}_m^v)^2 =  (-1)^m(\tilde{G}_x^v)^{2m} =  (-1)^m \left(T_{\bs a_1}^v\right)^m.\label{eq:specialMirror}
\ee
To exploit this, consider a cylinder whose `long' direction is parallel to ${\bs a}_2$ and is compact with a circumference of $m$ unit cells in the ${\bs a_1}$ direction, with $m$ odd. Periodic boundary conditions (b.c.) then require that $T_{\bs a_1}^m =1$. If the same property held true for single-vison states, namely that $\left(T_{\bs a_1}^v\right)^m = 1$, then it is evident that $\mathcal{M}_m$ exchanges edges and acts on a single-vison state with a nontrivial $-1$ factor. However, there is a subtlety: the vison could have anti-periodic boundary conditions, in which case $\left(T_{\bs a_1}^v\right)^m = -1$, and this generically will occur in two of the four MESs. (For instance, if we fix the vison to have periodic b.c. in the `$1$' MES, then it will also have periodic b.c. in the `$v
$' MES but anti-periodic b.c. in the `$b$' and `$f$' MESs.)
However, if we can {\it independently} determine $\left(T_{\bs a_1}^v\right)^m$, the nontrivial sign of $(\mathcal{M}_m^v)^2$  for $m$ odd can then be identified {\it relative} to the sign of  $\left(T_{\bs a_1}^v\right)^m$. We now discuss how to determine the vison boundary conditions.

Let us suppose that we start with the infinite-cylinder MES labeled by $a$.  Then, we can make a $T_{\bs a_1}$-symmetric entanglement cut, and look at the $T_{\bs a_1}$ eigenvalues of the Schmidt states.  Next, we can correspondingly do the same thing after threading a vison pair, i.e. in the $v\cdot a$ MES.  If the vison has anti-periodic boundary conditions, there should be a shift in the eigenvalues corresponding to $(T_{\bs a_1})^m = -1$; if it has periodic boundary conditions there should be no shift.  In other words, for the $a$ MES, all the $T_{\bs a_1}$ eigenvalues should become the same once we take them to the $m$th power, giving some constant, and similarly for the $v\cdot a$ MES.  These two constants could be the same (periodic vison b.c.) or differ by -1 (anti-periodic vison b.c.).

With this identification of the vison boundary conditions, we may  identify $\mathcal{M}_m^v $ as the `generalized mirror', whose quantum number fractionalization can be detected on odd-circumference cylinders.

 Finally, we turn to (\ref{eq:redanom3}). Using (\ref{eq:tildesGydefmod}) and (\ref{eq:sigmaxbarydefmod}), we may write this as
\bea
(T_{\bs a_1}^v)^{-1} (\tilde{G}_y^v)^{-1} &=& -\sigma^v_{x\bar{y}} \tilde{G}_x^v \sigma^v_{x\bar{y}}\nonumber\\
&=& -\left((\tilde{G}_x^v)^{-1} \sigma_{xy}^v \tilde{G}^v_x\right)\tilde{G}_x^v \left((\tilde{G}_x^v)^{-1} \sigma_{xy}^v \tilde{G}^v_x\right)\nonumber\\
&=& - (\tilde{G}_x^v)^{-1} \left(\sigma^v_{xy} \tilde{G}_x^v \sigma^v_{xy}\right) \tilde{G}_x^v\nonumber\\
&=& - (\tilde{G}_x^v)^{-1} \tilde{G}_y^v \tilde{G}_x^v.\label{eq:redanom3_manipulation_intermediate}
\eea
Now, using (\ref{eq:tildesGxGxTxmod}), we may rewrite (\ref{eq:redanom3_manipulation_intermediate}) solely in terms of $\tilde{G}^v_{x,y}$ and rearrange to arrive at the simple expression
\be
\left(\tilde{G}^v_y \tilde{G}^v_x\right)^2 = -1.
\ee
It is readily verified that under the action of this operation, the spatial coordinates are transformed as
\be
\tilde{G}_y \tilde{G}_x: (x,y) \mapsto (-x,-y) - (2,0),
\ee
which corresponds to inversion about a bond center at $(-1,0)$. So we have identified a third operator that exchanges edges on a finite cylinder and squares to (-1) acting on the single-vison sector. This completes our list of quantum numbers that are fractionalized for $\nu$ odd in the $p4g$ space group. 

\subsection{\label{subsec:experiments}Experimental Signatures}
We now briefly discuss the possibility of measuring space group symmetry fractionalization in experiments. Ref.~\onlinecite{essin2014spectroscopic} showed that the two-vison continuum  can have robust degeneracies that are a direct consequence of space group symmetry fractionalization, and that can be resolved by space group quantum numbers.  Such degeneracies can in principle be accessed via spectroscopic probes.

This phenomenon results from symmetry-protected degeneracies in the single-vison spectrum arising from the projective action of symmetry on visons. Such degeneracies occur when the vison transforms in a representation where the projective phase factors remain non-trivial under an enhanced $U(1)$ gauge freedom.  In other words, we must identify the set of relations in (\ref{eq:redanom1}-\ref{eq:redanom3}) that remain non-trivial even with this increase in gauge freedom (while of course, being careful not to introduce other non-trivial phase factors). Consider making a redefinition of the form $\tilde{G}_x^v \rightarrow \alpha \tilde{G}_x^v$, $\sigma_{xy}^v \rightarrow \beta \sigma_{xy}^v$, now with $\alpha, \beta\in U(1)$. Keeping the normal relations (\ref{eq:tildesGxGxTxmod} -\ref{eq:sigmasquaredmod}) fixed then requires 
 \bea
 T^v_{\bs a_1} &\rightarrow&  \alpha^2  T^v_{\bs a_1}, \nonumber\\
 \tilde{G}_y^v &\rightarrow&  \alpha \beta^2 \tilde{G}_y^v, \nonumber\\
  T^v_{\bs a_2} &\rightarrow&  \alpha^2  \alpha^2\beta^4 T^v_{\bs a_1}, \nonumber\\
  \sigma_{x\bar{y}}^v&\rightarrow& \beta \sigma_{x\bar{y}}^v,\nonumber\\
  \beta^2 &=& 1.
  \eea
 The last equation requires $\beta\in Z_2$, but $\alpha$ is still undetermined; turning to (\ref{eq:redanom1}-\ref{eq:redanom3}), we find that they are transformed to
\begin{subequations}
\bea
(\sigma_{xy}^v\sigma^v_{x\bar{y}})^2 &=& -1\label{eq:redanom1spect},\\
\tilde{G}^v_x T_{\bs a_2}^v &=& - \alpha^4(T_{\bs a_2}^v)^{-1} \tilde{G}^v_x\label{eq:redanom2spect},\\
(T_{\bs a_1}^v)^{-1} (\tilde{G}_y^v)^{-1} &=& -\alpha^4\sigma^v_{x\bar{y}} \tilde{G}_x^v \sigma^v_{x\bar{y}}\label{eq:redanom3spect}.
\eea
\end{subequations}
We see that relations (\ref{eq:redanom2spect}) and (\ref{eq:redanom3spect}) can be trivialized by utilizing the remaining gauge freedom to choose $\alpha^4 = -1$. Thus, the only remaining non-trivial relation is that associated with site centered inversion, $\sigma_{xy}^v\sigma_{x\bar{y}}^v$. We do not at present know of any simple way detect this fractionalization in experiments; we include this discussion merely to show that inversion symmetry is the {only} symmetry of $p4g$ whose fractionalization is  detectable {\it even in principle} by symmetry-protected degeneracies in the two-vison continuum.

\section{Concluding Remarks}
In this paper, we have studied the fractionalization of space-group quantum numbers by $Z_2$ topologically ordered phases in two dimensions, in situations where the ground state at integer filling is constrained to be non-trivial by extensions of the HOLSM theorem to non-symmorphic crystals. Specifically focusing on the $p4g$ space group, we have demonstrated that at odd integer filling, fractionalized phases of matter necessarily involve non-trivial anyon fluxes through each unit cell, resulting in the projective implementation of space group symmetries when acting on single-vison states. This in turn leads to robust numerical signatures of space group fractionalization, and a much more subtle possible spectroscopic signature in experiments. We have also demonstrated (Appendix~\ref{app:dorbitalglidefrac}) that our analysis is (to some extent) model-independent, suggesting that our symmetry considerations may be universal. 

In the future, it would be interesting to understand how to extend these ideas beyond the simple setting of this paper, and generalize the results reported here to higher dimensions and more complicated topological orders. We leave a discussion of such issues to future work.

\section*{Acknowledgements}
We are indebted to Mike Zaletel for several invariably enlightening discussions, and to Brett Brandom for collaboration on related work. We acknowledge support from NSF Grant DMR-1455366 (SAP), UC Irvine start-up funds (SAP, SBL) and KAIST start-up funds (SBL). The work of MH was supported by the U.S. Department of Energy, Office of Science, Basic Energy Sciences, under Award \# DE-FG-02-10ER46686 and subsequently under award \# DE-SC0014415.

\begin{appendix}
{

\section{Gauge-Fixing the Symmetry Relations \label{app:gaugefixing}}
In this section, we provide details on the gauge-fixing that leads to the relations (\ref{eq:tildesGxGxTxmod}-\ref{eq:sigmasquaredmod}) and (\ref{eq:redanom1}-\ref{eq:redanom3}). There is a $Z_2$ gauge ambiguity in defining the action of symmetries on single-vison states  that stems from the fact that operators always act on {\it physical} states that always contain a pair of anyons, and therefore we are free to introduce a (-1) into the action of an operator on any single anyon~\cite{PhysRevB.87.104406}. Before proceeding further, it is convenient to exploit this gauge freedom to `trivialize' as many of the symmetry relations as possible.

First, we translate both the  normal (\ref{eq:normalSGprodZ4_a}-\ref{eq:normalSGprodZ4_e}) and  non-trivial (\ref{eq:anomalousSGprodZ4_a}-\ref{eq:anomalousSGprodZ4_e}) symmetry relations for the single-vison states in terms of the redefined operators introduced in (\ref{eq:operatorredef}). We then have a new set of normal relations 
\begin{subequations}
\bea
(\tilde{G}^v_x)^2 &=&  T^v_{\bs a_1},\label{eq:tildesGxGxTx}\\
(\tilde{G}^v_y)^2 &=&  T^v_{\bs a_2},\label{eq:tildesGyGyTy}\\
\sigma_{x\bar{y}}^v &=& (\tilde{G}_x^v)^{-1} \sigma_{xy}^v \tilde{G}^v_x,\\
T^v_{\bs a_1}T^v_{\bs a_2}&=& T^v_{\bs a_2}T^v_{\bs a_1} ,\\
(\sigma^v_{xy})^2 &=& (\sigma^v_{x\bar{y}})^2 = 1,\\
(T_{\bs a_1}^v)^{-1} (\tilde{G}_y^v)^{-1} &=& \sigma^v_{x\bar{y}} \tilde{G}_x^v \sigma^v_{x\bar{y}}.
\eea
\end{subequations}
and (fewer) non-trivial ones,
\begin{subequations}
\bea
\tilde{G}_y^v &=& -\sigma_{xy}^v\tilde{G}_x^v \sigma_{xy}^v,\\
(\sigma_{xy}^v\sigma^v_{x\bar{y}})^2 &=& -1,\\
\tilde{G}^v_x T_{\bs a_2}^v &=& - (T_{\bs a_2}^v)^{-1} \tilde{G}^v_x.
\eea
\end{subequations}
We now utilize the $Z_2$ gauge freedom. As (\ref{eq:tildesGxGxTx}) and (\ref{eq:tildesGyGyTy}) each involve a single instance of the primitive translations, we cannot redefine $T_{\bs a_{1,2}}$ if we wish to keep these trivial. We can, however, transform $\tilde{G}_y \rightarrow -\tilde{G}_y$, in which case we have a modified set of normal relations
\begin{subequations}
\bea
(\tilde{G}^v_x)^2 &=&  T^v_{\bs a_1},\label{eq:tildesGxGxTxmodapp}\\
(\tilde{G}^v_y)^2 &=&  T^v_{\bs a_2},\label{eq:tildesGyGyTymodapp}\\
\tilde{G}_y^v &=& \sigma_{xy}^v\tilde{G}_x^v \sigma_{xy}^v,\label{eq:tildesGydefmodapp}\\\
\sigma_{x\bar{y}}^v &=& (\tilde{G}_x^v)^{-1} \sigma_{xy}^v \tilde{G}^v_x,\label{eq:sigmaxbarydefmodapp}\\
T^v_{\bs a_1}T^v_{\bs a_2}&=& T^v_{\bs a_2}T^v_{\bs a_1} ,\\
(\sigma^v_{xy})^2 &=& (\sigma^v_{x\bar{y}})^2 = 1\label{eq:sigmasquaredmodapp},
\eea
\end{subequations}
and nontrivial ones, 
\begin{subequations}
\bea
(\sigma_{xy}^v\sigma^v_{x\bar{y}})^2 &=& -1\label{eq:redanom1app},\\
\tilde{G}^v_x T_{\bs a_2}^v &=& - (T_{\bs a_2}^v)^{-1} \tilde{G}^v_x\label{eq:redanom2app},\\
(T_{\bs a_1}^v)^{-1} (\tilde{G}_y^v)^{-1} &=& -\sigma^v_{x\bar{y}} \tilde{G}_x^v \sigma^v_{x\bar{y}}\label{eq:redanom3app}.
\eea
\end{subequations}
Note that we have traded one nontrivial relation for another by this redefinition. The remaining gauge freedom is encapsulated by the transformations $\tilde{G}_{x,y}\rightarrow \alpha \tilde{G}_{x,y}$, $\sigma_{xy} \rightarrow \beta\sigma_{xy}$, $\sigma_{x\bar{y}} \rightarrow \beta\sigma_{x\bar{y}}$, where $\alpha, \beta\in Z_2$. It is readily verified that these factors cannot remove the non-trivial phase of $-1$  in (\ref{eq:redanom1}-\ref{eq:redanom3}), and hence the relations above characterize a nontrivial fractionalization class~\cite{PhysRevB.87.104406} (formally, an element of the cohomology group $H^2(p4g, Z_2)$.)  Therefore, in our discussion of identifying symmetry fractionalization in the main text, we focus exclusively on these three relations.

\section{Another $p4g$ example}
\label{app:dorbitalglidefrac}
In this appendix, we repeat our analysis in the main text for a different realization of $p4g$ symmetry in 2D. 
 \subsection{Checkerboard Lattice $d$-orbital Model and $Z_2$ gauge theory}
 The second lattice we study consists of $d$-orbitals arranged in a checkerboard pattern, so that the orbitals on the two checkerboard sublattices differ in their orientation by $\pi/2$: one  sublattice hosts $d_{(x+y)^2}$ orbitals, whereas the other hosts $d_{(x-y)^2}$ orbitals (see Fig.~~\ref{fig:checkerboard}). (The $d_{(x\pm y)^2}$ orbitals are identical in symmetry to the orbital conventionally denoted $d_{z^2}$ for an axis of quantization along $z$, but their quantization axes are chosen along  $(x\pm y)$.) Note that this model also has glide symmetries and belongs to space group $p4g$, but implements this in a distinct way from the SSL model. Here, the $d$-orbitals themselves transform nontrivially under the reflection portion of the glide, and swap their orientations; the subsequent half-translation is thus required to return the orbital arrangement to its original configuration on the checkerboard.  At $\nu=1$, the average site filling is $1/2$.

Lattice spin models with this underlying orbital structure may be of relevance to a class of perovskite materials that includes LaMnO$_3$ and KCuF$_3$. These materials form a three-dimensional bulk perovskite structure with low-temperature orbital and spin order. However, there is a dramatic difference in the relevant transition temperatures for the two orders;  for instance, the orbital ordering temperature $T_{OO} \sim 800$K while the N\'eel temperature $T_{N} \sim 140$K for LaMnO$_3$, and $T_{OO} \sim 800$K and $T_{N} \sim 40$K for KCuF$_3$. \cite{murakami1998resonant,lee2012two} This points to a large separation of spin and orbital ordering energy scales, suggesting that it is reasonable to consider the spin fluctuations against a background of frozen orbital order. A single layer of the resulting system would then be characterized by an effective Hamiltonian with the symmetries of $p4g$ encoded in its orbital pattern. We caution that at this point, it is not clear whether a simple model of the sort presented here will adequately describe the physics of any particular material candidate, but we merely provide this example to note that crystal symmetries can arise in a variety of ways especially in systems with orbital degrees of freedom. We note that a system of $\nu=1/2$ bosons on this lattice corresponds (via the hard-core mapping discussed above) to a $U(1)$-symmetric spin system with vanishing net magnetization, {\it i.e.} the $S^z_{\text{tot}} = 0$ sector of an XXZ spin model.

In Sec.~\ref{sec:Z4-gauge}, we studied the $Z_4$ gauge theory for $s$-orbital SSL model and investigated characteristics of low energy vison excitations. Based on a parton mean-field theory, we showed that glide reflections were fractionalized for odd integer unit cell fillings.
In this section, we repeat this analysis for the checkerboard-$d$-orbital lattice, to illustrate the universal aspects of the discussion in the main text.

We first derive the effective gauge theory of the checkerboard $d$-orbital model, focusing on $\nu=1$ (half-site-filling). We repeat the parton analysis choosing a translationally-invariant parton mean-field ansatz, with a single parton per site.
Accordingly, we define a parton number operator $n_{\bs r}$ with a conjugate phase variable $\phi_{\bs r}$ 
\bea
B_{\bs r}^\dagger &=& ( b_{\bs r}^\dagger)^2 \label{eq:partonz2map} \\
n_{\bs r} &=& 2 N_{\bs r}.  \label{eq:partonz2constraint}
\eea 
where $b_{\bs r}^\dagger=e^{i \phi_{\bs r}} $ is the parton creation operator. 
and (\ref{eq:partonz2constraint}) is the familiar constraint on the parton Hilbert space that reduces it to the physical Hilbert space of the original bosons.

Following the standard procedure of softening the constraint, we introduce a gauge field on each link of the lattice. In this case, the minimal choice for the gauge group is $Z_2$ and the Hilbert space on each link $\ell = ({ \bs r, \bs r' })$ consists of two states 
$ \mathfrak{h}_\ell = \{ | 0 \ra , | 1 \ra \}$. We introduce a $Z_2$ vector potential operator $a_l$ and the corresponding electric field operator $e_\ell$, which satisfy 
\bea
a_{\bs r \bs r'} | k \ra = e^{2\pi i k /2} | k \ra \text{  and  } e_{\bs r \bs r'} | k \ra = | k+1 \ra . 
\label{eq:linkactionZ2}
\eea 
Note that as we work in a $Z_2$ theory we may identify $| 2 \ra $ and $| 0 \ra$; also, we have that $a_{\bs r' \bs r} =  a^\dagger_{\bs r \bs r'} = a_{\bs r \bs r'}$, and similarly for $e_{\bs r' \bs r}$.

In terms of gauge fields, the constraint (\ref{eq:partonz2constraint}) can be represented as a Gauss law for the $Z_2$ electric field,  \bea
\underset{ \bs r' \in \la \bs r \bs r' \ra }{\prod} e_{\bs r \bs r'} = e^{ i \pi  n_{\bs r} },
\label{eq:z2gausslaw}
\eea
where the product is over the links connecting ${\bs r}$ to its neighboring sites ${\bs r}'$.
 
In parallel with the analysis in the main text, we rewrite the microscopic Bose-Hubbard model  (\ref{eq:bosonHamBase}) by placing the partons in a gapped phase and integrating them out, leaving an effective gauge theory defined on square plaquettes,  
\be
{{\mathcal H}_g} &=& - h \sum_{\la \bs r \bs r' \ra} ( e_{\bs r \bs r'} + \text{h.c.}) - K \sum_{p\in \Box} (\underset{ \bs r \bs r'\in p}{  \prod} a_{\bs r \bs r'} + \text{h.c.} ), \nn \\  
\label{eq:z2gaugeHam}
\ee
where $h,K >0$.  

For $h \gg K$, the the ground state of (\ref{eq:z2gaugeHam}) has $e_{\bs r \bs r'} = 1$   
and we have $n_{\bs r} \equiv 0 \text{ (mod 2) }$ from the Gauss law (\ref{eq:z2gausslaw}). This is the electric-field-confined phase, where the partons are bound into the original bosons, and (following the HOLSM theorem) the ground state must break symmetry if it is gapped but not fractionalized. 

For $K \gg h$, the partons can propagate independently. 
This corresponds to the deconfined phase, where the Gauss law (\ref{eq:z2gausslaw}) is represented as 
\be
\underset{ \bs r' \in \la \bs r \bs r' \ra }{\prod} e_{\bs r \bs r'} = -1.
\label{eq:frozenz2gauss}
\ee
 Fig.~\ref{fig:checkerboard} shows an electric field satisfying (\ref{eq:frozenz2gauss}). We emphasize that (as before) the $Z_2$ electric field configuration does not enlarge the unit cell, so it does not break translation symmetry. However, it {\it does} break point group symmetry, in particular,  the glide symmetry. We now turn to an analysis of the low-energy vison excitations of this gauge theory.

\begin{figure}[b]
{\includegraphics[width=.7\columnwidth]{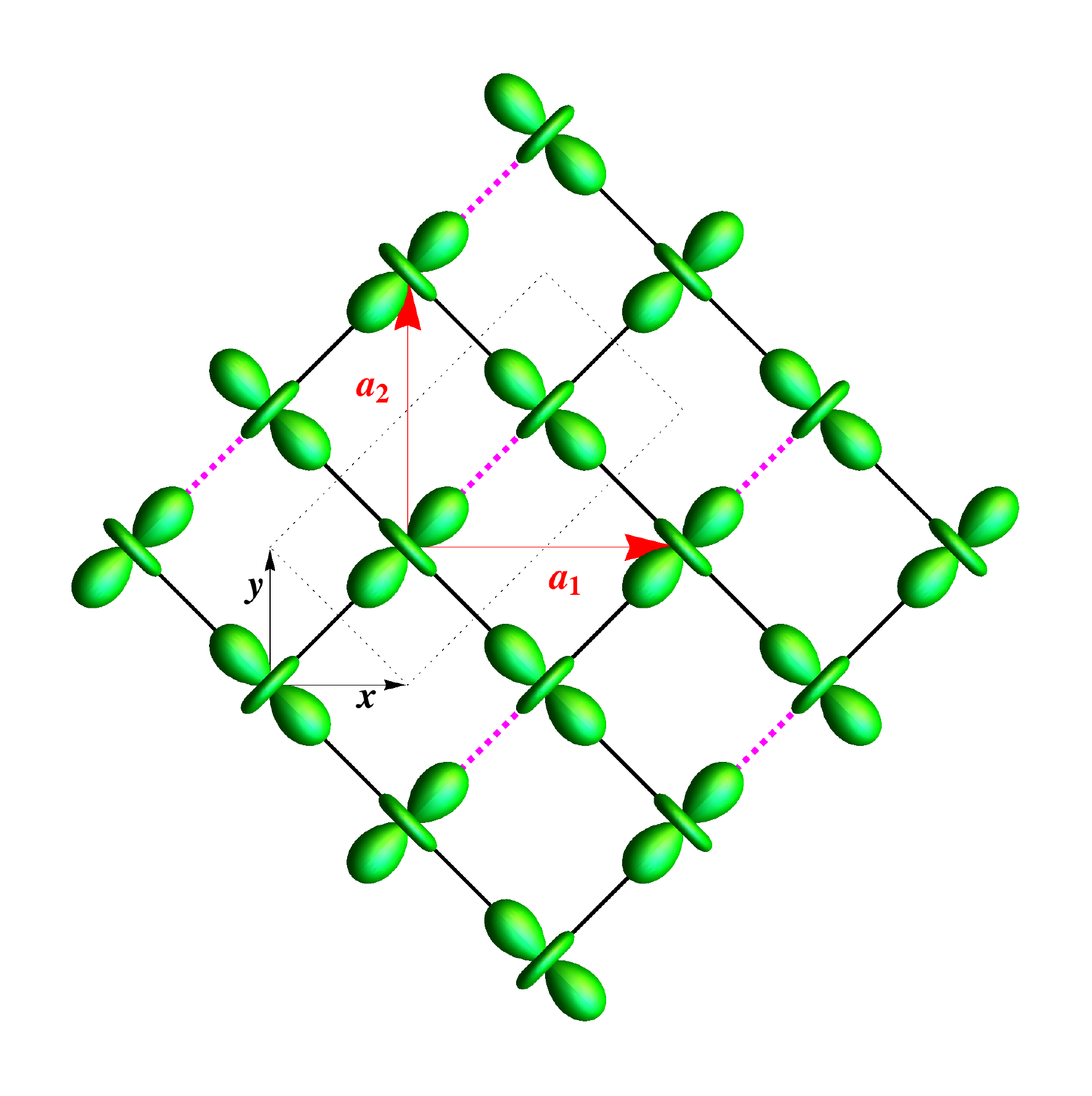}}
\caption{(color online) Schematic picture of $d$-orbital ordering pattern between $d_{(x+y)^2 }$ and $d_{(x-y)^2}$ on $x$-$y$ plane. Dashed line indicates the enlarged unit cell in the presence of orbital ordering. Solid (Black colored) and dashed (magenta colored) links have electric field configuration $e_{\bs r \bs r'} =  1$ and $-1$ respectively, satisfying $\underset{ \bs r' \in \la \bs r \bs r' \ra }{\prod} e_{\bs r \bs r'} =-1$. (See the main text for details.)  }
\label{fig:checkerboard}
\end{figure}


\subsection{Dual Ising  model}
\label{subsec:dual-ising}
In order to investigate the properties of visons in the $Z_2$ gauge theory, it is once again convenient to study the dual theory, here a quantum Ising model.   
The dual lattice is also a checkerboard lattice with two sites in each unit cell. 
At each dual lattice site ${\bsbr}$, we introduce a new set of Ising operators $E_\bsbr $ and $A_\bsbr$ that satisfy (\ref{eq:z4dualA}-\ref{eq:z4dualE}) in terms of which the dual Ising theory then can be written exactly as (\ref{eq:Z4clock}), 
as for the dual $Z_4$ clock model in the main text. 
Under the duality, the Gauss law constraint is mapped to a constraint on the product of bonds $\eta_{\bsbr \bsbr'}$ for every dual checkerboard lattice plaquette: we have
\be
\underset{\bsbr \bsbr' \in \Box}{\prod} \eta_{\bsbr \bsbr'} = (-1)^\nu.
\label{eq:frustratedbondZ2ising}
\ee
at boson filling $\nu$. 
For $\nu=1$, the product of bond variables satisfies -1 and the dual theory is the {\it fully frustrated Ising model} on the checkerboard lattice:
each plaquette has exactly one frustrated bond with $\eta_{\bsbr \bsbr'} =-1$. As a reference configuration (Fig.~\ref{fig:checkerboard}), we may assign the links $ \ell =( \bsbr \bsbr')$ on the dual checkerboard lattice that bisect the dashed magenta links of the direct lattice to have $\eta_{\bsbr \bsbr'} = -1$, while the ones that bisect the solid black links have $\eta_{\bsbr \bsbr'} = 1$. 

In the dual  theory, the non-local vison creation operator is described by a local operator $A_{\bsbr}$, along with a (static) non-local string product of bond strengths $\eta_{\bsbr \bsbr'}$,  fixed by the reference configuration above. We may now examine its transformation properties under various lattice symmetries.

\subsection{Vison symmetry analysis}

Before we consider the transformation of $A_{\bsbr}$, we first list the symmetries of the checkerboard $d$-orbital model.
There are two translations along the lattice vectors ($\bs a_1 = (2,0)$ and $\bs a_2 = (0,2)$): 
\bea
T_{\bs a_1} &:& (x,y) \rightarrow (x,y) + \bs a_1
\nn \\
T_{\bs a_2} &:& (x,y) \rightarrow (x,y) + \bs a_2 
\label{eq:checker-trans} 
\eea
In addition, the lattice is symmetric under  mirror reflections with axes rotated from the lattice vectors by $\pi/4$: 
\bea
\sigma_{xy} &:& (x,y) \rightarrow (y, x)
\nn \\
\sigma_{x\bar{y}} &:& (x,y) \rightarrow (-y,-x) 
\label{eq:checker-mirror} 
\eea
and under glide reflections for axes parallel to the lattice vectors: 
\bea
G_{x} &:& (x,y) \rightarrow (x,-y) + \bs a_1 /2
\nn  \\ 
G_{y} &:& (x,y) \rightarrow (-x,y) + \bs a_2 /2.
\label{eq:checker-glide} 
\eea
Fig.~\ref{fig:checkerboard} shows the lattice structure and lattice vectors. 
We note that these symmetries are identical to those of the $s$-orbital SSL model, underscoring the fact that both lattices have $p4g$ symmetry. 
We now proceed to list the transformation properties of the  vison operator $A_{\bsbr}$  under (\ref{eq:checker-trans}-\ref{eq:checker-glide}).
For clarity, we denote the   vison  annihilation operator at sublattice $s \in \{1,2\}$ and unit cell position $(m \bs a_1 + n \bs a_2 )$ by $A_{(m,n)  s}$. 
\begin{table*}
\centering
\begin{tabular}{| p{2cm}||p{3cm} | p{3cm} | p{3cm} | p{3cm}  |} 
\hline 
&$ \sigma_{x\bar{y}} $ & $ \sigma_{xy}$ & $G_{x}$ & $G_{y}$  \\
\hline \hline 
$A_{(m,n) 1} $ & $ (-1)^{m+n} A^\dagger_{(-n, -m) 2} $ & $A^\dagger_{(n-1,m)2 }$ 
& $(-1)^{n+1} A^\dagger_{(m,-n) 2}$ & $(-1)^m A^\dagger_{(-m-1,n) 2}$ \\
\hline
$ A_{(m,n) 2}$ & $(-1)^{m+n} A^\dagger_{(-n,-m) 1}$ & $A^\dagger_{(n,m+1) 1}$ 
& $ (-1)^n A^\dagger_{ (m+1,-n) 1} $ & $ (-1)^m A^\dagger_{(-m-1,n+1) 1}$ \\
\hline
\end{tabular}
\caption{Projective symmetry group : Transformation of vison operator $A_{\bsbr}$ under four lattice symmetries, for a given phase configuration $\eta_{\bsbr \bsbr'} $ that satisfies $\underset{\bsbr \bsbr' \in \square}{\prod} \eta_{\bsbr \bsbr'} = -1$. (see Fig.~\ref{fig:checkerboard})}
\label{tab:visonSymmetryZ2}
\end{table*}
Table.~\ref{tab:visonSymmetryZ2} lists the transformation properties of a single vison state $A_{\bsbr}^\dagger | 0\ra $ under the different symmetries.

Under a subset of the space group symmetries, the single-vison state transforms `trivially': 
\begin{subequations}
\bea
T^v_{\bs a_1} T^v_{\bs a_2}  &=& T^v_{\bs a_2} T^v_{\bs a_1}
\label{eq:normalSGprodZ2_a} \\
(\sigma_{xy}^v)^2  &=& 1
\label{eq:normalSGprodZ2_b} \\
(\sigma_{x\bar{y}}^v)^2 &=& 1
\label{eq:normalSGprodZ2_c} \\
G_y^v  \sigma_{xy}^v  &=& \sigma_{xy}^v G_x^v
\label{eq:normalSGprodZ2_d} \\
T_x^v (G_y^v)^{-1} &=& \sigma_{x\bar{y}}^v G_x \sigma_{x\bar{y}} ^v
\label{eq:normalSGprodZ2_e} 
\eea
\end{subequations}The remaining space group combinations involve non-trivial phase factors of (-1): 
\begin{subequations}
\bea
(G_x^v)^2 &=& -T^v_{\bs a_1} 
\label{eq:anomalousSGprodZ2_a}\\ 
(G_y^v)^2 &=& -T^v_{\bs a_2} 
\label{eq:anomalousSGprodZ2_b}\\ 
\sigma^v_{x\bar{y}} G_x^v &=& - G_x^v \sigma_{xy}^v 
\label{eq:anomalousSGprodZ2_c}\\ 
\sigma^v_{x\bar{y}} \sigma_{xy}^v &=& - \sigma_{xy}^v \sigma_{x\bar{y}}^v 
\label{eq:anomalousSGprodZ2_d}\\ 
G_x^v T_{\bs a_2}^v &=& - (T^v_{\bs a_2})^{-1} G_x^v. 
\label{eq:anomalousSGprodZ2_e}
\eea
\end{subequations}
These phase factors (\ref{eq:anomalousSGprodZ2_a}-\ref{eq:anomalousSGprodZ2_e}) once again reflect the fact that the visons fractionalize symmetry.  Note that the entire set of algebraic relationships is identical to those [(\ref{eq:anomalousSGprodZ4_a}-\ref{eq:anomalousSGprodZ4_e})] computed in the $Z_4$ case in the main text.
It is straightforward to verify that the nontrivial phase factors are  absent fo $\nu=2$, in consonance with the HOLSM argument that requires no fractionalization at this filling.

\subsection{Condensing visons: broken-symmetry phases}
\label{subsec:z2-subsec:vison-condensation}
Having studied the symmetries of the single vison state in the deconfined phase of the pure $Z_2$ gauge theory, 
we now briefly discuss the  proximate phases accessed by condensing a single vison. 
For the case of $Z_2$ gauge theory, the charges $e$ and fluxes $m$ take values $q_e,q_m \in \{ 0, 1 \}$ and the mutual statistics phase factor is given by $e^{i \pi q_e q_m}$ when an $e$ particle is taken around an $m$ particle (or vice-versa).  
Confined phases may be accessed by condensing a single vison $\la A_{(m,n)\mu} \ra  \neq 0$; at $\nu=1$, the nontrivial phase factor leads to broken glide symmetry, as anticipated from the HOLSM theorem.

\subsection{Relation between $Z_4$ and $Z_2$ theories}
 The analysis in this section underscores the universal nature of the arguments presented here: even though the $Z_4$ theory is apparently richer, as long as we restrict our attention to $\nu=1$, both models admit no trivial confined phases, while permitting a deconfined phase with $Z_2$ topological order. Put differently, while the $Z_4$ gauge structure is a possibility, there is no {\it symmetry} that enforces it, as all the HOLSM constraints are satisfied by the simpler $Z_2$ theory that descends from it.
}
\end{appendix}

\bibliography{frac-glide-2d-bib}

\begin{thebibliography}{59}%
\makeatletter
\providecommand \@ifxundefined [1]{%
 \@ifx{#1\undefined}
}%
\providecommand \@ifnum [1]{%
 \ifnum #1\expandafter \@firstoftwo
 \else \expandafter \@secondoftwo
 \fi
}%
\providecommand \@ifx [1]{%
 \ifx #1\expandafter \@firstoftwo
 \else \expandafter \@secondoftwo
 \fi
}%
\providecommand \natexlab [1]{#1}%
\providecommand \enquote  [1]{``#1''}%
\providecommand \bibnamefont  [1]{#1}%
\providecommand \bibfnamefont [1]{#1}%
\providecommand \citenamefont [1]{#1}%
\providecommand \href@noop [0]{\@secondoftwo}%
\providecommand \href [0]{\begingroup \@sanitize@url \@href}%
\providecommand \@href[1]{\@@startlink{#1}\@@href}%
\providecommand \@@href[1]{\endgroup#1\@@endlink}%
\providecommand \@sanitize@url [0]{\catcode `\\12\catcode `\$12\catcode
  `\&12\catcode `\#12\catcode `\^12\catcode `\_12\catcode `\%12\relax}%
\providecommand \@@startlink[1]{}%
\providecommand \@@endlink[0]{}%
\providecommand \url  [0]{\begingroup\@sanitize@url \@url }%
\providecommand \@url [1]{\endgroup\@href {#1}{\urlprefix }}%
\providecommand \urlprefix  [0]{URL }%
\providecommand \Eprint [0]{\href }%
\providecommand \doibase [0]{http://dx.doi.org/}%
\providecommand \selectlanguage [0]{\@gobble}%
\providecommand \bibinfo  [0]{\@secondoftwo}%
\providecommand \bibfield  [0]{\@secondoftwo}%
\providecommand \translation [1]{[#1]}%
\providecommand \BibitemOpen [0]{}%
\providecommand \bibitemStop [0]{}%
\providecommand \bibitemNoStop [0]{.\EOS\space}%
\providecommand \EOS [0]{\spacefactor3000\relax}%
\providecommand \BibitemShut  [1]{\csname bibitem#1\endcsname}%
\let\auto@bib@innerbib\@empty
\bibitem [{\citenamefont {Yan}\ \emph {et~al.}(2011)\citenamefont {Yan},
  \citenamefont {Huse},\ and\ \citenamefont {White}}]{yan2011spin}%
  \BibitemOpen
  \bibfield  {author} {\bibinfo {author} {\bibfnamefont {S.}~\bibnamefont
  {Yan}}, \bibinfo {author} {\bibfnamefont {D.~A.}\ \bibnamefont {Huse}}, \
  and\ \bibinfo {author} {\bibfnamefont {S.~R.}\ \bibnamefont {White}},\ }\href
  {http://www.sciencemag.org/content/332/6034/1173.short} {\bibfield  {journal}
  {\bibinfo  {journal} {Science}\ }\textbf {\bibinfo {volume} {332}},\ \bibinfo
  {pages} {1173} (\bibinfo {year} {2011})}\BibitemShut {NoStop}%
\bibitem [{\citenamefont {Wen}(2002)}]{WenPSG}%
  \BibitemOpen
  \bibfield  {author} {\bibinfo {author} {\bibfnamefont {X.-G.}\ \bibnamefont
  {Wen}},\ }\href {\doibase 10.1103/PhysRevB.65.165113} {\bibfield  {journal}
  {\bibinfo  {journal} {Phys. Rev. B}\ }\textbf {\bibinfo {volume} {65}},\
  \bibinfo {pages} {165113} (\bibinfo {year} {2002})}\BibitemShut {NoStop}%
\bibitem [{\citenamefont {Essin}\ and\ \citenamefont
  {Hermele}(2013)}]{PhysRevB.87.104406}%
  \BibitemOpen
  \bibfield  {author} {\bibinfo {author} {\bibfnamefont {A.~M.}\ \bibnamefont
  {Essin}}\ and\ \bibinfo {author} {\bibfnamefont {M.}~\bibnamefont
  {Hermele}},\ }\href {\doibase 10.1103/PhysRevB.87.104406} {\bibfield
  {journal} {\bibinfo  {journal} {Phys. Rev. B}\ }\textbf {\bibinfo {volume}
  {87}},\ \bibinfo {pages} {104406} (\bibinfo {year} {2013})}\BibitemShut
  {NoStop}%
\bibitem [{\citenamefont {Mesaros}\ and\ \citenamefont
  {Ran}(2013)}]{PhysRevB.87.155115}%
  \BibitemOpen
  \bibfield  {author} {\bibinfo {author} {\bibfnamefont {A.}~\bibnamefont
  {Mesaros}}\ and\ \bibinfo {author} {\bibfnamefont {Y.}~\bibnamefont {Ran}},\
  }\href {\doibase 10.1103/PhysRevB.87.155115} {\bibfield  {journal} {\bibinfo
  {journal} {Phys. Rev. B}\ }\textbf {\bibinfo {volume} {87}},\ \bibinfo
  {pages} {155115} (\bibinfo {year} {2013})}\BibitemShut {NoStop}%
\bibitem [{\citenamefont {Hung}\ and\ \citenamefont {Wen}(2013)}]{HungWen}%
  \BibitemOpen
  \bibfield  {author} {\bibinfo {author} {\bibfnamefont {L.-Y.}\ \bibnamefont
  {Hung}}\ and\ \bibinfo {author} {\bibfnamefont {X.-G.}\ \bibnamefont {Wen}},\
  }\href {\doibase 10.1103/PhysRevB.87.165107} {\bibfield  {journal} {\bibinfo
  {journal} {Phys. Rev. B}\ }\textbf {\bibinfo {volume} {87}},\ \bibinfo
  {pages} {165107} (\bibinfo {year} {2013})}\BibitemShut {NoStop}%
\bibitem [{\citenamefont {Lieb}\ \emph {et~al.}(1961)\citenamefont {Lieb},
  \citenamefont {Schultz},\ and\ \citenamefont {Mattis}}]{Lieb1961407}%
  \BibitemOpen
  \bibfield  {author} {\bibinfo {author} {\bibfnamefont {E.}~\bibnamefont
  {Lieb}}, \bibinfo {author} {\bibfnamefont {T.}~\bibnamefont {Schultz}}, \
  and\ \bibinfo {author} {\bibfnamefont {D.}~\bibnamefont {Mattis}},\ }\href
  {\doibase http://dx.doi.org/10.1016/0003-4916(61)90115-4} {\bibfield
  {journal} {\bibinfo  {journal} {Annals of Physics}\ }\textbf {\bibinfo
  {volume} {16}},\ \bibinfo {pages} {407 } (\bibinfo {year}
  {1961})}\BibitemShut {NoStop}%
\bibitem [{\citenamefont {Oshikawa}(2000)}]{oshikawa2000topological}%
  \BibitemOpen
  \bibfield  {author} {\bibinfo {author} {\bibfnamefont {M.}~\bibnamefont
  {Oshikawa}},\ }\href
  {http://journals.aps.org/prl/abstract/10.1103/PhysRevLett.84.3370} {\bibfield
   {journal} {\bibinfo  {journal} {Physical Review Letters}\ }\textbf {\bibinfo
  {volume} {84}},\ \bibinfo {pages} {3370} (\bibinfo {year}
  {2000})}\BibitemShut {NoStop}%
\bibitem [{\citenamefont {Hastings}(2004)}]{PhysRevB.69.104431}%
  \BibitemOpen
  \bibfield  {author} {\bibinfo {author} {\bibfnamefont {M.~B.}\ \bibnamefont
  {Hastings}},\ }\href {\doibase 10.1103/PhysRevB.69.104431} {\bibfield
  {journal} {\bibinfo  {journal} {Phys. Rev. B}\ }\textbf {\bibinfo {volume}
  {69}},\ \bibinfo {pages} {104431} (\bibinfo {year} {2004})}\BibitemShut
  {NoStop}%
\bibitem [{\citenamefont {Hastings}(2005)}]{hastings2005sufficient}%
  \BibitemOpen
  \bibfield  {author} {\bibinfo {author} {\bibfnamefont {M.~B.}\ \bibnamefont
  {Hastings}},\ }\href {http://stacks.iop.org/0295-5075/70/i=6/a=824}
  {\bibfield  {journal} {\bibinfo  {journal} {EPL (Europhysics Letters)}\
  }\textbf {\bibinfo {volume} {70}},\ \bibinfo {pages} {824} (\bibinfo {year}
  {2005})}\BibitemShut {NoStop}%
\bibitem [{\citenamefont {Paramekanti}\ and\ \citenamefont
  {Vishwanath}(2004)}]{paramekanti2004extending}%
  \BibitemOpen
  \bibfield  {author} {\bibinfo {author} {\bibfnamefont {A.}~\bibnamefont
  {Paramekanti}}\ and\ \bibinfo {author} {\bibfnamefont {A.}~\bibnamefont
  {Vishwanath}},\ }\href
  {http://journals.aps.org/prb/abstract/10.1103/PhysRevB.70.245118} {\bibfield
  {journal} {\bibinfo  {journal} {Physical Review B}\ }\textbf {\bibinfo
  {volume} {70}},\ \bibinfo {pages} {245118} (\bibinfo {year}
  {2004})}\BibitemShut {NoStop}%
\bibitem [{\citenamefont {Essin}\ and\ \citenamefont
  {Hermele}(2014)}]{essin2014spectroscopic}%
  \BibitemOpen
  \bibfield  {author} {\bibinfo {author} {\bibfnamefont {A.~M.}\ \bibnamefont
  {Essin}}\ and\ \bibinfo {author} {\bibfnamefont {M.}~\bibnamefont
  {Hermele}},\ }\href
  {http://journals.aps.org/prb/abstract/10.1103/PhysRevB.90.121102} {\bibfield
  {journal} {\bibinfo  {journal} {Physical Review B}\ }\textbf {\bibinfo
  {volume} {90}},\ \bibinfo {pages} {121102} (\bibinfo {year}
  {2014})}\BibitemShut {NoStop}%
\bibitem [{\citenamefont {{Cheng}}\ \emph {et~al.}(2015)\citenamefont
  {{Cheng}}, \citenamefont {{Zaletel}}, \citenamefont {{Barkeshli}},
  \citenamefont {{Vishwanath}},\ and\ \citenamefont
  {{Bonderson}}}]{StationQSurfaceHOLSM}%
  \BibitemOpen
  \bibfield  {author} {\bibinfo {author} {\bibfnamefont {M.}~\bibnamefont
  {{Cheng}}}, \bibinfo {author} {\bibfnamefont {M.}~\bibnamefont {{Zaletel}}},
  \bibinfo {author} {\bibfnamefont {M.}~\bibnamefont {{Barkeshli}}}, \bibinfo
  {author} {\bibfnamefont {A.}~\bibnamefont {{Vishwanath}}}, \ and\ \bibinfo
  {author} {\bibfnamefont {P.}~\bibnamefont {{Bonderson}}},\ }\href@noop {}
  {\bibfield  {journal} {\bibinfo  {journal} {ArXiv e-prints}\ } (\bibinfo
  {year} {2015})},\ \Eprint {http://arxiv.org/abs/1511.02263} {arXiv:1511.02263
  [cond-mat.str-el]} \BibitemShut {NoStop}%
\bibitem [{\citenamefont {Parameswaran}\ \emph {et~al.}(2013)\citenamefont
  {Parameswaran}, \citenamefont {Turner}, \citenamefont {Arovas},\ and\
  \citenamefont {Vishwanath}}]{parameswaran2013topological}%
  \BibitemOpen
  \bibfield  {author} {\bibinfo {author} {\bibfnamefont {S.~A.}\ \bibnamefont
  {Parameswaran}}, \bibinfo {author} {\bibfnamefont {A.~M.}\ \bibnamefont
  {Turner}}, \bibinfo {author} {\bibfnamefont {D.~P.}\ \bibnamefont {Arovas}},
  \ and\ \bibinfo {author} {\bibfnamefont {A.}~\bibnamefont {Vishwanath}},\
  }\href {http://www.nature.com/nphys/journal/v9/n5/full/nphys2600.html}
  {\bibfield  {journal} {\bibinfo  {journal} {Nature Physics}\ }\textbf
  {\bibinfo {volume} {9}},\ \bibinfo {pages} {299} (\bibinfo {year}
  {2013})}\BibitemShut {NoStop}%
\bibitem [{Note1()}]{Note1}%
  \BibitemOpen
  \bibinfo {note} {In $d=3$, the large number of non-symmorphic space groups
  stems from the additional possibility of {\protect \it screw rotations}, a
  rotation combined with a fractional lattice translation parallel to the
  rotation axis.}\BibitemShut {Stop}%
\bibitem [{\citenamefont {Liu}\ \emph {et~al.}(2014)\citenamefont {Liu},
  \citenamefont {Zhang},\ and\ \citenamefont {VanLeeuwen}}]{TNCI_PRB_Original}%
  \BibitemOpen
  \bibfield  {author} {\bibinfo {author} {\bibfnamefont {C.-X.}\ \bibnamefont
  {Liu}}, \bibinfo {author} {\bibfnamefont {R.-X.}\ \bibnamefont {Zhang}}, \
  and\ \bibinfo {author} {\bibfnamefont {B.~K.}\ \bibnamefont {VanLeeuwen}},\
  }\href {\doibase 10.1103/PhysRevB.90.085304} {\bibfield  {journal} {\bibinfo
  {journal} {Phys. Rev. B}\ }\textbf {\bibinfo {volume} {90}},\ \bibinfo
  {pages} {085304} (\bibinfo {year} {2014})}\BibitemShut {NoStop}%
\bibitem [{\citenamefont {Fang}\ and\ \citenamefont {Fu}(2015)}]{FangFu}%
  \BibitemOpen
  \bibfield  {author} {\bibinfo {author} {\bibfnamefont {C.}~\bibnamefont
  {Fang}}\ and\ \bibinfo {author} {\bibfnamefont {L.}~\bibnamefont {Fu}},\
  }\href {\doibase 10.1103/PhysRevB.91.161105} {\bibfield  {journal} {\bibinfo
  {journal} {Phys. Rev. B}\ }\textbf {\bibinfo {volume} {91}},\ \bibinfo
  {pages} {161105} (\bibinfo {year} {2015})}\BibitemShut {NoStop}%
\bibitem [{\citenamefont {Wang}\ \emph {et~al.}(2016)\citenamefont {Wang},
  \citenamefont {Alexandradinata}, \citenamefont {Cava},\ and\ \citenamefont
  {Bernevig}}]{HourglassFermions}%
  \BibitemOpen
  \bibfield  {author} {\bibinfo {author} {\bibfnamefont {Z.}~\bibnamefont
  {Wang}}, \bibinfo {author} {\bibfnamefont {A.}~\bibnamefont
  {Alexandradinata}}, \bibinfo {author} {\bibfnamefont {R.~J.}\ \bibnamefont
  {Cava}}, \ and\ \bibinfo {author} {\bibfnamefont {B.~A.}\ \bibnamefont
  {Bernevig}},\ }\href {http://dx.doi.org/10.1038/nature17410} {\bibfield
  {journal} {\bibinfo  {journal} {Nature}\ }\textbf {\bibinfo {volume} {532}},\
  \bibinfo {pages} {189} (\bibinfo {year} {2016})}\BibitemShut {NoStop}%
\bibitem [{\citenamefont {Alexandradinata}\ \emph {et~al.}(2016)\citenamefont
  {Alexandradinata}, \citenamefont {Wang},\ and\ \citenamefont
  {Bernevig}}]{CohomologicalInsulators}%
  \BibitemOpen
  \bibfield  {author} {\bibinfo {author} {\bibfnamefont {A.}~\bibnamefont
  {Alexandradinata}}, \bibinfo {author} {\bibfnamefont {Z.}~\bibnamefont
  {Wang}}, \ and\ \bibinfo {author} {\bibfnamefont {B.~A.}\ \bibnamefont
  {Bernevig}},\ }\href {\doibase 10.1103/PhysRevX.6.021008} {\bibfield
  {journal} {\bibinfo  {journal} {Phys. Rev. X}\ }\textbf {\bibinfo {volume}
  {6}},\ \bibinfo {pages} {021008} (\bibinfo {year} {2016})}\BibitemShut
  {NoStop}%
\bibitem [{\citenamefont {Shiozaki}\ \emph {et~al.}(2016)\citenamefont
  {Shiozaki}, \citenamefont {Sato},\ and\ \citenamefont
  {Gomi}}]{TNCI_Classification}%
  \BibitemOpen
  \bibfield  {author} {\bibinfo {author} {\bibfnamefont {K.}~\bibnamefont
  {Shiozaki}}, \bibinfo {author} {\bibfnamefont {M.}~\bibnamefont {Sato}}, \
  and\ \bibinfo {author} {\bibfnamefont {K.}~\bibnamefont {Gomi}},\ }\href
  {\doibase 10.1103/PhysRevB.93.195413} {\bibfield  {journal} {\bibinfo
  {journal} {Phys. Rev. B}\ }\textbf {\bibinfo {volume} {93}},\ \bibinfo
  {pages} {195413} (\bibinfo {year} {2016})}\BibitemShut {NoStop}%
\bibitem [{\citenamefont {{Hermele}}\ and\ \citenamefont
  {{Chen}}(2015)}]{HermeleChenFFAnomaly}%
  \BibitemOpen
  \bibfield  {author} {\bibinfo {author} {\bibfnamefont {M.}~\bibnamefont
  {{Hermele}}}\ and\ \bibinfo {author} {\bibfnamefont {X.}~\bibnamefont
  {{Chen}}},\ }\href@noop {} {\bibfield  {journal} {\bibinfo  {journal} {ArXiv
  e-prints}\ } (\bibinfo {year} {2015})},\ \Eprint
  {http://arxiv.org/abs/1508.00573} {arXiv:1508.00573 [cond-mat.str-el]}
  \BibitemShut {NoStop}%
\bibitem [{\citenamefont {Shastry}\ and\ \citenamefont
  {Sutherland}(1981)}]{shastry1981exact}%
  \BibitemOpen
  \bibfield  {author} {\bibinfo {author} {\bibfnamefont {B.~S.}\ \bibnamefont
  {Shastry}}\ and\ \bibinfo {author} {\bibfnamefont {B.}~\bibnamefont
  {Sutherland}},\ }\href
  {http://www.sciencedirect.com/science/article/pii/037843638190838X}
  {\bibfield  {journal} {\bibinfo  {journal} {Physica B+ C}\ }\textbf {\bibinfo
  {volume} {108}},\ \bibinfo {pages} {1069} (\bibinfo {year}
  {1981})}\BibitemShut {NoStop}%
\bibitem [{\citenamefont {Sebastian}\ \emph {et~al.}(2008)\citenamefont
  {Sebastian}, \citenamefont {Harrison}, \citenamefont {Sengupta},
  \citenamefont {Batista}, \citenamefont {Francoual}, \citenamefont {Palm},
  \citenamefont {Murphy}, \citenamefont {Marcano}, \citenamefont {Dabkowska},\
  and\ \citenamefont {Gaulin}}]{sebastian2008fractalization}%
  \BibitemOpen
  \bibfield  {author} {\bibinfo {author} {\bibfnamefont {S.~E.}\ \bibnamefont
  {Sebastian}}, \bibinfo {author} {\bibfnamefont {N.}~\bibnamefont {Harrison}},
  \bibinfo {author} {\bibfnamefont {P.}~\bibnamefont {Sengupta}}, \bibinfo
  {author} {\bibfnamefont {C.}~\bibnamefont {Batista}}, \bibinfo {author}
  {\bibfnamefont {S.}~\bibnamefont {Francoual}}, \bibinfo {author}
  {\bibfnamefont {E.}~\bibnamefont {Palm}}, \bibinfo {author} {\bibfnamefont
  {T.}~\bibnamefont {Murphy}}, \bibinfo {author} {\bibfnamefont
  {N.}~\bibnamefont {Marcano}}, \bibinfo {author} {\bibfnamefont
  {H.}~\bibnamefont {Dabkowska}}, \ and\ \bibinfo {author} {\bibfnamefont
  {B.}~\bibnamefont {Gaulin}},\ }\href
  {http://www.pnas.org/content/105/51/20157.short} {\bibfield  {journal}
  {\bibinfo  {journal} {Proceedings of the National Academy of Sciences}\
  }\textbf {\bibinfo {volume} {105}},\ \bibinfo {pages} {20157} (\bibinfo
  {year} {2008})}\BibitemShut {NoStop}%
\bibitem [{\citenamefont {Kim}\ \emph {et~al.}(2008)\citenamefont {Kim},
  \citenamefont {Bennett},\ and\ \citenamefont {Aronson}}]{PhysRevB.77.144425}%
  \BibitemOpen
  \bibfield  {author} {\bibinfo {author} {\bibfnamefont {M.~S.}\ \bibnamefont
  {Kim}}, \bibinfo {author} {\bibfnamefont {M.~C.}\ \bibnamefont {Bennett}}, \
  and\ \bibinfo {author} {\bibfnamefont {M.~C.}\ \bibnamefont {Aronson}},\
  }\href {\doibase 10.1103/PhysRevB.77.144425} {\bibfield  {journal} {\bibinfo
  {journal} {Phys. Rev. B}\ }\textbf {\bibinfo {volume} {77}},\ \bibinfo
  {pages} {144425} (\bibinfo {year} {2008})}\BibitemShut {NoStop}%
\bibitem [{\citenamefont {Michimura}\ \emph {et~al.}(2006)\citenamefont
  {Michimura}, \citenamefont {Shigekawa}, \citenamefont {Iga}, \citenamefont
  {Sera}, \citenamefont {Takabatake}, \citenamefont {Ohoyama},\ and\
  \citenamefont {Okabe}}]{michimura2006magnetic}%
  \BibitemOpen
  \bibfield  {author} {\bibinfo {author} {\bibfnamefont {S.}~\bibnamefont
  {Michimura}}, \bibinfo {author} {\bibfnamefont {A.}~\bibnamefont
  {Shigekawa}}, \bibinfo {author} {\bibfnamefont {F.}~\bibnamefont {Iga}},
  \bibinfo {author} {\bibfnamefont {M.}~\bibnamefont {Sera}}, \bibinfo {author}
  {\bibfnamefont {T.}~\bibnamefont {Takabatake}}, \bibinfo {author}
  {\bibfnamefont {K.}~\bibnamefont {Ohoyama}}, \ and\ \bibinfo {author}
  {\bibfnamefont {Y.}~\bibnamefont {Okabe}},\ }\href
  {http://www.sciencedirect.com/science/article/pii/S0921452606002857}
  {\bibfield  {journal} {\bibinfo  {journal} {Physica B: Condensed Matter}\
  }\textbf {\bibinfo {volume} {378}},\ \bibinfo {pages} {596} (\bibinfo {year}
  {2006})}\BibitemShut {NoStop}%
\bibitem [{\citenamefont {Yoshii}\ \emph {et~al.}(2006)\citenamefont {Yoshii},
  \citenamefont {Yamamoto}, \citenamefont {Hagiwara}, \citenamefont
  {Shigekawa}, \citenamefont {Michimura}, \citenamefont {Iga}, \citenamefont
  {Takabatake},\ and\ \citenamefont {Kindo}}]{1742-6596-51-1-011}%
  \BibitemOpen
  \bibfield  {author} {\bibinfo {author} {\bibfnamefont {S.}~\bibnamefont
  {Yoshii}}, \bibinfo {author} {\bibfnamefont {T.}~\bibnamefont {Yamamoto}},
  \bibinfo {author} {\bibfnamefont {M.}~\bibnamefont {Hagiwara}}, \bibinfo
  {author} {\bibfnamefont {A.}~\bibnamefont {Shigekawa}}, \bibinfo {author}
  {\bibfnamefont {S.}~\bibnamefont {Michimura}}, \bibinfo {author}
  {\bibfnamefont {F.}~\bibnamefont {Iga}}, \bibinfo {author} {\bibfnamefont
  {T.}~\bibnamefont {Takabatake}}, \ and\ \bibinfo {author} {\bibfnamefont
  {K.}~\bibnamefont {Kindo}},\ }\href
  {http://stacks.iop.org/1742-6596/51/i=1/a=011} {\bibfield  {journal}
  {\bibinfo  {journal} {Journal of Physics: Conference Series}\ }\textbf
  {\bibinfo {volume} {51}},\ \bibinfo {pages} {59} (\bibinfo {year}
  {2006})}\BibitemShut {NoStop}%
\bibitem [{\citenamefont {Jaime}\ \emph {et~al.}(2012)\citenamefont {Jaime},
  \citenamefont {Daou}, \citenamefont {Crooker}, \citenamefont {Weickert},
  \citenamefont {Uchida}, \citenamefont {Feiguin}, \citenamefont {Batista},
  \citenamefont {Dabkowska},\ and\ \citenamefont
  {Gaulin}}]{jaime2012magnetostriction}%
  \BibitemOpen
  \bibfield  {author} {\bibinfo {author} {\bibfnamefont {M.}~\bibnamefont
  {Jaime}}, \bibinfo {author} {\bibfnamefont {R.}~\bibnamefont {Daou}},
  \bibinfo {author} {\bibfnamefont {S.~A.}\ \bibnamefont {Crooker}}, \bibinfo
  {author} {\bibfnamefont {F.}~\bibnamefont {Weickert}}, \bibinfo {author}
  {\bibfnamefont {A.}~\bibnamefont {Uchida}}, \bibinfo {author} {\bibfnamefont
  {A.~E.}\ \bibnamefont {Feiguin}}, \bibinfo {author} {\bibfnamefont {C.~D.}\
  \bibnamefont {Batista}}, \bibinfo {author} {\bibfnamefont {H.~A.}\
  \bibnamefont {Dabkowska}}, \ and\ \bibinfo {author} {\bibfnamefont {B.~D.}\
  \bibnamefont {Gaulin}},\ }\href
  {http://www.pnas.org/content/109/31/12404.short} {\bibfield  {journal}
  {\bibinfo  {journal} {Proceedings of the National Academy of Sciences}\
  }\textbf {\bibinfo {volume} {109}},\ \bibinfo {pages} {12404} (\bibinfo
  {year} {2012})}\BibitemShut {NoStop}%
\bibitem [{\citenamefont {Mat'a{\v s}}\ \emph {et~al.}(2010)\citenamefont
  {Mat'a{\v s}}, \citenamefont {Siemensmeyer}, \citenamefont {Wheeler},
  \citenamefont {Wulf}, \citenamefont {Beyer}, \citenamefont
  {Hermannsd{\"o}rfer}, \citenamefont {Ignatchik}, \citenamefont {Uhlarz},
  \citenamefont {Flachbart}, \citenamefont {Gab{\'a}ni}, \citenamefont
  {Priputen}, \citenamefont {Efdokimova},\ and\ \citenamefont
  {Shitsevalova}}]{1742-6596-200-3-032041}%
  \BibitemOpen
  \bibfield  {author} {\bibinfo {author} {\bibfnamefont {S.}~\bibnamefont
  {Mat'a{\v s}}}, \bibinfo {author} {\bibfnamefont {K.}~\bibnamefont
  {Siemensmeyer}}, \bibinfo {author} {\bibfnamefont {E.}~\bibnamefont
  {Wheeler}}, \bibinfo {author} {\bibfnamefont {E.}~\bibnamefont {Wulf}},
  \bibinfo {author} {\bibfnamefont {R.}~\bibnamefont {Beyer}}, \bibinfo
  {author} {\bibfnamefont {T.}~\bibnamefont {Hermannsd{\"o}rfer}}, \bibinfo
  {author} {\bibfnamefont {O.}~\bibnamefont {Ignatchik}}, \bibinfo {author}
  {\bibfnamefont {M.}~\bibnamefont {Uhlarz}}, \bibinfo {author} {\bibfnamefont
  {K.}~\bibnamefont {Flachbart}}, \bibinfo {author} {\bibfnamefont
  {S.}~\bibnamefont {Gab{\'a}ni}}, \bibinfo {author} {\bibfnamefont
  {P.}~\bibnamefont {Priputen}}, \bibinfo {author} {\bibfnamefont
  {A.}~\bibnamefont {Efdokimova}}, \ and\ \bibinfo {author} {\bibfnamefont
  {N.}~\bibnamefont {Shitsevalova}},\ }\href
  {http://stacks.iop.org/1742-6596/200/i=3/a=032041} {\bibfield  {journal}
  {\bibinfo  {journal} {Journal of Physics: Conference Series}\ }\textbf
  {\bibinfo {volume} {200}},\ \bibinfo {pages} {032041} (\bibinfo {year}
  {2010})}\BibitemShut {NoStop}%
\bibitem [{\citenamefont {Matsuda}\ \emph {et~al.}(2013)\citenamefont
  {Matsuda}, \citenamefont {Abe}, \citenamefont {Takeyama}, \citenamefont
  {Kageyama}, \citenamefont {Corboz}, \citenamefont {Honecker}, \citenamefont
  {Manmana}, \citenamefont {Foltin}, \citenamefont {Schmidt},\ and\
  \citenamefont {Mila}}]{matsuda2013magnetization}%
  \BibitemOpen
  \bibfield  {author} {\bibinfo {author} {\bibfnamefont {Y.}~\bibnamefont
  {Matsuda}}, \bibinfo {author} {\bibfnamefont {N.}~\bibnamefont {Abe}},
  \bibinfo {author} {\bibfnamefont {S.}~\bibnamefont {Takeyama}}, \bibinfo
  {author} {\bibfnamefont {H.}~\bibnamefont {Kageyama}}, \bibinfo {author}
  {\bibfnamefont {P.}~\bibnamefont {Corboz}}, \bibinfo {author} {\bibfnamefont
  {A.}~\bibnamefont {Honecker}}, \bibinfo {author} {\bibfnamefont
  {S.}~\bibnamefont {Manmana}}, \bibinfo {author} {\bibfnamefont
  {G.}~\bibnamefont {Foltin}}, \bibinfo {author} {\bibfnamefont
  {K.}~\bibnamefont {Schmidt}}, \ and\ \bibinfo {author} {\bibfnamefont
  {F.}~\bibnamefont {Mila}},\ }\href@noop {} {\bibfield  {journal} {\bibinfo
  {journal} {Physical review letters}\ }\textbf {\bibinfo {volume} {111}},\
  \bibinfo {pages} {137204} (\bibinfo {year} {2013})}\BibitemShut {NoStop}%
\bibitem [{\citenamefont {{Barkeshli}}\ \emph {et~al.}(2014)\citenamefont
  {{Barkeshli}}, \citenamefont {{Bonderson}}, \citenamefont {{Cheng}},\ and\
  \citenamefont {{Wang}}}]{StationQ_Gcrossed}%
  \BibitemOpen
  \bibfield  {author} {\bibinfo {author} {\bibfnamefont {M.}~\bibnamefont
  {{Barkeshli}}}, \bibinfo {author} {\bibfnamefont {P.}~\bibnamefont
  {{Bonderson}}}, \bibinfo {author} {\bibfnamefont {M.}~\bibnamefont
  {{Cheng}}}, \ and\ \bibinfo {author} {\bibfnamefont {Z.}~\bibnamefont
  {{Wang}}},\ }\href@noop {} {\bibfield  {journal} {\bibinfo  {journal} {ArXiv
  e-prints}\ } (\bibinfo {year} {2014})},\ \Eprint
  {http://arxiv.org/abs/1410.4540} {arXiv:1410.4540 [cond-mat.str-el]}
  \BibitemShut {NoStop}%
\bibitem [{\citenamefont {{Qi}}\ \emph {et~al.}(2015)\citenamefont {{Qi}},
  \citenamefont {{Cheng}},\ and\ \citenamefont {{Fang}}}]{QiChengFangVisons}%
  \BibitemOpen
  \bibfield  {author} {\bibinfo {author} {\bibfnamefont {Y.}~\bibnamefont
  {{Qi}}}, \bibinfo {author} {\bibfnamefont {M.}~\bibnamefont {{Cheng}}}, \
  and\ \bibinfo {author} {\bibfnamefont {C.}~\bibnamefont {{Fang}}},\
  }\href@noop {} {\bibfield  {journal} {\bibinfo  {journal} {ArXiv e-prints}\ }
  (\bibinfo {year} {2015})},\ \Eprint {http://arxiv.org/abs/1509.02927}
  {arXiv:1509.02927 [cond-mat.str-el]} \BibitemShut {NoStop}%
\bibitem [{\citenamefont {Kogut}(1979)}]{kogut1979introduction}%
  \BibitemOpen
  \bibfield  {author} {\bibinfo {author} {\bibfnamefont {J.~B.}\ \bibnamefont
  {Kogut}},\ }\href
  {http://journals.aps.org/rmp/abstract/10.1103/RevModPhys.51.659} {\bibfield
  {journal} {\bibinfo  {journal} {Reviews of Modern Physics}\ }\textbf
  {\bibinfo {volume} {51}},\ \bibinfo {pages} {659} (\bibinfo {year}
  {1979})}\BibitemShut {NoStop}%
\bibitem [{\citenamefont {Rothe}(2012)}]{rothe2012lattice}%
  \BibitemOpen
  \bibfield  {author} {\bibinfo {author} {\bibfnamefont {H.~J.}\ \bibnamefont
  {Rothe}},\ }\href@noop {} {\emph {\bibinfo {title} {Lattice gauge
  theories}}}\ (\bibinfo  {publisher} {World Scientific},\ \bibinfo {year}
  {2012})\BibitemShut {NoStop}%
\bibitem [{\citenamefont {Altman}\ and\ \citenamefont
  {Auerbach}(1998)}]{altman1998haldane}%
  \BibitemOpen
  \bibfield  {author} {\bibinfo {author} {\bibfnamefont {E.}~\bibnamefont
  {Altman}}\ and\ \bibinfo {author} {\bibfnamefont {A.}~\bibnamefont
  {Auerbach}},\ }\href
  {http://journals.aps.org/prl/abstract/10.1103/PhysRevLett.81.4484} {\bibfield
   {journal} {\bibinfo  {journal} {Physical review letters}\ }\textbf {\bibinfo
  {volume} {81}},\ \bibinfo {pages} {4484} (\bibinfo {year}
  {1998})}\BibitemShut {NoStop}%
\bibitem [{\citenamefont {Oshikawa}\ \emph {et~al.}(1997)\citenamefont
  {Oshikawa}, \citenamefont {Yamanaka},\ and\ \citenamefont
  {Affleck}}]{oshikawa1997magnetization}%
  \BibitemOpen
  \bibfield  {author} {\bibinfo {author} {\bibfnamefont {M.}~\bibnamefont
  {Oshikawa}}, \bibinfo {author} {\bibfnamefont {M.}~\bibnamefont {Yamanaka}},
  \ and\ \bibinfo {author} {\bibfnamefont {I.}~\bibnamefont {Affleck}},\ }\href
  {http://journals.aps.org/prl/abstract/10.1103/PhysRevLett.78.1984} {\bibfield
   {journal} {\bibinfo  {journal} {Physical review letters}\ }\textbf {\bibinfo
  {volume} {78}},\ \bibinfo {pages} {1984} (\bibinfo {year}
  {1997})}\BibitemShut {NoStop}%
\bibitem [{\citenamefont {Oshikawa}(2003)}]{oshikawa2003insulator}%
  \BibitemOpen
  \bibfield  {author} {\bibinfo {author} {\bibfnamefont {M.}~\bibnamefont
  {Oshikawa}},\ }\href {\doibase 10.1103/PhysRevLett.90.236401} {\bibfield
  {journal} {\bibinfo  {journal} {Phys. Rev. Lett.}\ }\textbf {\bibinfo
  {volume} {90}},\ \bibinfo {pages} {236401} (\bibinfo {year}
  {2003})}\BibitemShut {NoStop}%
\bibitem [{Note2()}]{Note2}%
  \BibitemOpen
  \bibinfo {note} {Although rigorous energy bounds can only be given for a
  different -- but gauge-equivalent -- flux insertion, for pedagogical reasons
  we keep a simpler choice with the understanding that the arguments of
  Ref.~\protect \rev@citealpnum {hastings2005sufficient} can be applied to make
  our proofs rigorous.}\BibitemShut {Stop}%
\bibitem [{Note3()}]{Note3}%
  \BibitemOpen
  \bibinfo {note} {We can generalize the results to any lattice and to $d=3$,
  but for the cases studied in this paper, $d=2$ and square lattice symmetry is
  sufficient}\BibitemShut {NoStop}%
\bibitem [{\citenamefont {K{\"o}nig}\ and\ \citenamefont
  {Mermin}(1997)}]{konig1997electronic}%
  \BibitemOpen
  \bibfield  {author} {\bibinfo {author} {\bibfnamefont {A.}~\bibnamefont
  {K{\"o}nig}}\ and\ \bibinfo {author} {\bibfnamefont {N.~D.}\ \bibnamefont
  {Mermin}},\ }\href
  {http://journals.aps.org/prb/abstract/10.1103/PhysRevB.56.13607} {\bibfield
  {journal} {\bibinfo  {journal} {Physical Review B}\ }\textbf {\bibinfo
  {volume} {56}},\ \bibinfo {pages} {13607} (\bibinfo {year}
  {1997})}\BibitemShut {NoStop}%
\bibitem [{\citenamefont {Watanabe}\ \emph {et~al.}(2015)\citenamefont
  {Watanabe}, \citenamefont {Po}, \citenamefont {Vishwanath},\ and\
  \citenamefont {Zaletel}}]{Watanabe24112015}%
  \BibitemOpen
  \bibfield  {author} {\bibinfo {author} {\bibfnamefont {H.}~\bibnamefont
  {Watanabe}}, \bibinfo {author} {\bibfnamefont {H.~C.}\ \bibnamefont {Po}},
  \bibinfo {author} {\bibfnamefont {A.}~\bibnamefont {Vishwanath}}, \ and\
  \bibinfo {author} {\bibfnamefont {M.}~\bibnamefont {Zaletel}},\ }\href
  {\doibase 10.1073/pnas.1514665112} {\bibfield  {journal} {\bibinfo  {journal}
  {Proceedings of the National Academy of Sciences}\ }\textbf {\bibinfo
  {volume} {112}},\ \bibinfo {pages} {14551} (\bibinfo {year} {2015})},\
  \Eprint
  {http://arxiv.org/abs/http://www.pnas.org/content/112/47/14551.full.pdf}
  {http://www.pnas.org/content/112/47/14551.full.pdf} \BibitemShut {NoStop}%
\bibitem [{\citenamefont {Balents}\ \emph {et~al.}(2002)\citenamefont
  {Balents}, \citenamefont {Fisher},\ and\ \citenamefont {Girvin}}]{balents02}%
  \BibitemOpen
  \bibfield  {author} {\bibinfo {author} {\bibfnamefont {L.}~\bibnamefont
  {Balents}}, \bibinfo {author} {\bibfnamefont {M.~P.~A.}\ \bibnamefont
  {Fisher}}, \ and\ \bibinfo {author} {\bibfnamefont {S.~M.}\ \bibnamefont
  {Girvin}},\ }\href {\doibase 10.1103/PhysRevB.65.224412} {\bibfield
  {journal} {\bibinfo  {journal} {Phys. Rev. B}\ }\textbf {\bibinfo {volume}
  {65}},\ \bibinfo {pages} {224412} (\bibinfo {year} {2002})},\ \Eprint
  {http://arxiv.org/abs/cond-mat/0110005} {arXiv:cond-mat/0110005} \BibitemShut
  {NoStop}%
\bibitem [{\citenamefont {Balents}\ \emph {et~al.}(1999)\citenamefont
  {Balents}, \citenamefont {Fisher},\ and\ \citenamefont {Nayak}}]{balents99}%
  \BibitemOpen
  \bibfield  {author} {\bibinfo {author} {\bibfnamefont {L.}~\bibnamefont
  {Balents}}, \bibinfo {author} {\bibfnamefont {M.~P.~A.}\ \bibnamefont
  {Fisher}}, \ and\ \bibinfo {author} {\bibfnamefont {C.}~\bibnamefont
  {Nayak}},\ }\href {\doibase 10.1103/PhysRevB.60.1654} {\bibfield  {journal}
  {\bibinfo  {journal} {Phys. Rev. B}\ }\textbf {\bibinfo {volume} {60}},\
  \bibinfo {pages} {1654} (\bibinfo {year} {1999})},\ \Eprint
  {http://arxiv.org/abs/cond-mat/9811236} {arXiv:cond-mat/9811236} \BibitemShut
  {NoStop}%
\bibitem [{\citenamefont {Kitaev}(2003)}]{kitaev03}%
  \BibitemOpen
  \bibfield  {author} {\bibinfo {author} {\bibfnamefont {A.~{\relax Yu}.}\
  \bibnamefont {Kitaev}},\ }\href {\doibase 10.1016/S0003-4916(02)00018-0}
  {\bibfield  {journal} {\bibinfo  {journal} {Ann. Phys.}\ }\textbf {\bibinfo
  {volume} {303}},\ \bibinfo {pages} {2} (\bibinfo {year} {2003})},\ \Eprint
  {http://arxiv.org/abs/quant-ph/9707021} {arXiv:quant-ph/9707021} \BibitemShut
  {NoStop}%
\bibitem [{\citenamefont {Moessner}\ and\ \citenamefont
  {Sondhi}(2001)}]{moessner01a}%
  \BibitemOpen
  \bibfield  {author} {\bibinfo {author} {\bibfnamefont {R.}~\bibnamefont
  {Moessner}}\ and\ \bibinfo {author} {\bibfnamefont {S.~L.}\ \bibnamefont
  {Sondhi}},\ }\href {\doibase 10.1103/PhysRevLett.86.1881} {\bibfield
  {journal} {\bibinfo  {journal} {Phys. Rev. Lett.}\ }\textbf {\bibinfo
  {volume} {86}},\ \bibinfo {pages} {1881} (\bibinfo {year} {2001})},\ \Eprint
  {http://arxiv.org/abs/cond-mat/0007378} {arXiv:cond-mat/0007378} \BibitemShut
  {NoStop}%
\bibitem [{\citenamefont {Moessner}\ \emph {et~al.}(2001)\citenamefont
  {Moessner}, \citenamefont {Sondhi},\ and\ \citenamefont
  {Fradkin}}]{moessner01b}%
  \BibitemOpen
  \bibfield  {author} {\bibinfo {author} {\bibfnamefont {R.}~\bibnamefont
  {Moessner}}, \bibinfo {author} {\bibfnamefont {S.~L.}\ \bibnamefont
  {Sondhi}}, \ and\ \bibinfo {author} {\bibfnamefont {E.}~\bibnamefont
  {Fradkin}},\ }\href {\doibase 10.1103/PhysRevB.65.024504} {\bibfield
  {journal} {\bibinfo  {journal} {Phys. Rev. B}\ }\textbf {\bibinfo {volume}
  {65}},\ \bibinfo {pages} {024504} (\bibinfo {year} {2001})},\ \Eprint
  {http://arxiv.org/abs/cond-mat/0103396} {arXiv:cond-mat/0103396} \BibitemShut
  {NoStop}%
\bibitem [{\citenamefont {Read}\ and\ \citenamefont {Sachdev}(1991)}]{read91}%
  \BibitemOpen
  \bibfield  {author} {\bibinfo {author} {\bibfnamefont {N.}~\bibnamefont
  {Read}}\ and\ \bibinfo {author} {\bibfnamefont {S.}~\bibnamefont {Sachdev}},\
  }\href {\doibase 10.1103/PhysRevLett.66.1773} {\bibfield  {journal} {\bibinfo
   {journal} {Phys. Rev. Lett.}\ }\textbf {\bibinfo {volume} {66}},\ \bibinfo
  {pages} {1773} (\bibinfo {year} {1991})}\BibitemShut {NoStop}%
\bibitem [{\citenamefont {Read}\ and\ \citenamefont
  {Chakraborty}(1989)}]{chakraborty89}%
  \BibitemOpen
  \bibfield  {author} {\bibinfo {author} {\bibfnamefont {N.}~\bibnamefont
  {Read}}\ and\ \bibinfo {author} {\bibfnamefont {B.}~\bibnamefont
  {Chakraborty}},\ }\href {\doibase 10.1103/PhysRevB.40.7133} {\bibfield
  {journal} {\bibinfo  {journal} {Phys. Rev. B}\ }\textbf {\bibinfo {volume}
  {40}},\ \bibinfo {pages} {7133} (\bibinfo {year} {1989})}\BibitemShut
  {NoStop}%
\bibitem [{\citenamefont {Senthil}\ and\ \citenamefont
  {Fisher}(2000)}]{senthil00}%
  \BibitemOpen
  \bibfield  {author} {\bibinfo {author} {\bibfnamefont {T.}~\bibnamefont
  {Senthil}}\ and\ \bibinfo {author} {\bibfnamefont {M.~P.~A.}\ \bibnamefont
  {Fisher}},\ }\href {\doibase 10.1103/PhysRevB.62.7850} {\bibfield  {journal}
  {\bibinfo  {journal} {Phys. Rev. B}\ }\textbf {\bibinfo {volume} {62}},\
  \bibinfo {pages} {7850} (\bibinfo {year} {2000})},\ \Eprint
  {http://arxiv.org/abs/cond-mat/9910224} {arXiv:cond-mat/9910224} \BibitemShut
  {NoStop}%
\bibitem [{\citenamefont {Wen}(1991)}]{wen91}%
  \BibitemOpen
  \bibfield  {author} {\bibinfo {author} {\bibfnamefont {X.~G.}\ \bibnamefont
  {Wen}},\ }\href {\doibase 10.1103/PhysRevB.44.2664} {\bibfield  {journal}
  {\bibinfo  {journal} {Phys. Rev. B}\ }\textbf {\bibinfo {volume} {44}},\
  \bibinfo {pages} {2664} (\bibinfo {year} {1991})}\BibitemShut {NoStop}%
\bibitem [{\citenamefont {Senthil}\ and\ \citenamefont
  {Fisher}(2001)}]{SenthilFisher}%
  \BibitemOpen
  \bibfield  {author} {\bibinfo {author} {\bibfnamefont {T.}~\bibnamefont
  {Senthil}}\ and\ \bibinfo {author} {\bibfnamefont {M.~P.~A.}\ \bibnamefont
  {Fisher}},\ }\href {\doibase 10.1103/PhysRevB.63.134521} {\bibfield
  {journal} {\bibinfo  {journal} {Phys. Rev. B}\ }\textbf {\bibinfo {volume}
  {63}},\ \bibinfo {pages} {134521} (\bibinfo {year} {2001})}\BibitemShut
  {NoStop}%
\bibitem [{\citenamefont {Kageyama}\ \emph {et~al.}(1999)\citenamefont
  {Kageyama}, \citenamefont {Yoshimura}, \citenamefont {Stern}, \citenamefont
  {Mushnikov}, \citenamefont {Onizuka}, \citenamefont {Kato}, \citenamefont
  {Kosuge}, \citenamefont {Slichter}, \citenamefont {Goto},\ and\ \citenamefont
  {Ueda}}]{kageyama1999exact}%
  \BibitemOpen
  \bibfield  {author} {\bibinfo {author} {\bibfnamefont {H.}~\bibnamefont
  {Kageyama}}, \bibinfo {author} {\bibfnamefont {K.}~\bibnamefont {Yoshimura}},
  \bibinfo {author} {\bibfnamefont {R.}~\bibnamefont {Stern}}, \bibinfo
  {author} {\bibfnamefont {N.~V.}\ \bibnamefont {Mushnikov}}, \bibinfo {author}
  {\bibfnamefont {K.}~\bibnamefont {Onizuka}}, \bibinfo {author} {\bibfnamefont
  {M.}~\bibnamefont {Kato}}, \bibinfo {author} {\bibfnamefont {K.}~\bibnamefont
  {Kosuge}}, \bibinfo {author} {\bibfnamefont {C.~P.}\ \bibnamefont
  {Slichter}}, \bibinfo {author} {\bibfnamefont {T.}~\bibnamefont {Goto}}, \
  and\ \bibinfo {author} {\bibfnamefont {Y.}~\bibnamefont {Ueda}},\ }\href
  {\doibase 10.1103/PhysRevLett.82.3168} {\bibfield  {journal} {\bibinfo
  {journal} {Phys. Rev. Lett.}\ }\textbf {\bibinfo {volume} {82}},\ \bibinfo
  {pages} {3168} (\bibinfo {year} {1999})}\BibitemShut {NoStop}%
\bibitem [{Note4()}]{Note4}%
  \BibitemOpen
  \bibinfo {note} {Note that the term `vison' is conventionally associated with
  Ising gauge theories; we use the term here more generally to describe fluxes
  in discrete gauge theories.}\BibitemShut {Stop}%
\bibitem [{Note5()}]{Note5}%
  \BibitemOpen
  \bibinfo {note} {Note that we may verify that the flux pattern preserves
  symmetry for $\nu =2$.}\BibitemShut {Stop}%
\bibitem [{\citenamefont {K\"onig}\ and\ \citenamefont
  {Mermin}(1999)}]{KonigMerminPNAS}%
  \BibitemOpen
  \bibfield  {author} {\bibinfo {author} {\bibfnamefont {A.}~\bibnamefont
  {K\"onig}}\ and\ \bibinfo {author} {\bibfnamefont {N.~D.}\ \bibnamefont
  {Mermin}},\ }\href@noop {} {\bibfield  {journal} {\bibinfo  {journal} {Proc.
  Natl. Acad. Sci. USA}\ }\textbf {\bibinfo {volume} {96}},\ \bibinfo {pages}
  {3502} (\bibinfo {year} {1999})}\BibitemShut {NoStop}%
\bibitem [{Note6()}]{Note6}%
  \BibitemOpen
  \bibinfo {note} {To see this, first observe that combining particles with
  $q_m=1$ and $q_m=3$ in a $Z_4$ gauge theory should be equivalent to the
  vacuum, but since the $q_m=3$ particle is condensed and is hence equivalent
  to the vacuum, it follows that the $q_m=1$ particle must also be condensed. A
  similar argument based on the fact that combining $q_m=3$ and $q_m=2$ should
  give $q_m=1$ reveals that $q_m=2$ is also condensed.}\BibitemShut {Stop}%
\bibitem [{\citenamefont {{Zaletel}}\ \emph {et~al.}(2015)\citenamefont
  {{Zaletel}}, \citenamefont {{Lu}},\ and\ \citenamefont
  {{Vishwanath}}}]{ZaletelPSGMeasurement}%
  \BibitemOpen
  \bibfield  {author} {\bibinfo {author} {\bibfnamefont {M.}~\bibnamefont
  {{Zaletel}}}, \bibinfo {author} {\bibfnamefont {Y.-M.}\ \bibnamefont {{Lu}}},
  \ and\ \bibinfo {author} {\bibfnamefont {A.}~\bibnamefont {{Vishwanath}}},\
  }\href@noop {} {\bibfield  {journal} {\bibinfo  {journal} {ArXiv e-prints}\ }
  (\bibinfo {year} {2015})},\ \Eprint {http://arxiv.org/abs/1501.01395}
  {arXiv:1501.01395 [cond-mat.str-el]} \BibitemShut {NoStop}%
\bibitem [{\citenamefont {Qi}\ and\ \citenamefont {Fu}(2015)}]{QiFu}%
  \BibitemOpen
  \bibfield  {author} {\bibinfo {author} {\bibfnamefont {Y.}~\bibnamefont
  {Qi}}\ and\ \bibinfo {author} {\bibfnamefont {L.}~\bibnamefont {Fu}},\ }\href
  {\doibase 10.1103/PhysRevB.91.100401} {\bibfield  {journal} {\bibinfo
  {journal} {Phys. Rev. B}\ }\textbf {\bibinfo {volume} {91}},\ \bibinfo
  {pages} {100401} (\bibinfo {year} {2015})}\BibitemShut {NoStop}%
\bibitem [{Note7()}]{Note7}%
  \BibitemOpen
  \bibinfo {note} {Since we focus on $Z_2$ phases, where each particle is its
  own antiparticle, we omit the bar on antiparticles henceforth.}\BibitemShut
  {Stop}%
\bibitem [{\citenamefont {Murakami}\ \emph {et~al.}(1998)\citenamefont
  {Murakami}, \citenamefont {Hill}, \citenamefont {Gibbs}, \citenamefont
  {Blume}, \citenamefont {Koyama}, \citenamefont {Tanaka}, \citenamefont
  {Kawata}, \citenamefont {Arima}, \citenamefont {Tokura}, \citenamefont
  {Hirota},\ and\ \citenamefont {Endoh}}]{murakami1998resonant}%
  \BibitemOpen
  \bibfield  {author} {\bibinfo {author} {\bibfnamefont {Y.}~\bibnamefont
  {Murakami}}, \bibinfo {author} {\bibfnamefont {J.~P.}\ \bibnamefont {Hill}},
  \bibinfo {author} {\bibfnamefont {D.}~\bibnamefont {Gibbs}}, \bibinfo
  {author} {\bibfnamefont {M.}~\bibnamefont {Blume}}, \bibinfo {author}
  {\bibfnamefont {I.}~\bibnamefont {Koyama}}, \bibinfo {author} {\bibfnamefont
  {M.}~\bibnamefont {Tanaka}}, \bibinfo {author} {\bibfnamefont
  {H.}~\bibnamefont {Kawata}}, \bibinfo {author} {\bibfnamefont
  {T.}~\bibnamefont {Arima}}, \bibinfo {author} {\bibfnamefont
  {Y.}~\bibnamefont {Tokura}}, \bibinfo {author} {\bibfnamefont
  {K.}~\bibnamefont {Hirota}}, \ and\ \bibinfo {author} {\bibfnamefont
  {Y.}~\bibnamefont {Endoh}},\ }\href {\doibase 10.1103/PhysRevLett.81.582}
  {\bibfield  {journal} {\bibinfo  {journal} {Phys. Rev. Lett.}\ }\textbf
  {\bibinfo {volume} {81}},\ \bibinfo {pages} {582} (\bibinfo {year}
  {1998})}\BibitemShut {NoStop}%
\bibitem [{\citenamefont {Lee}\ \emph {et~al.}(2012)\citenamefont {Lee},
  \citenamefont {Yuan}, \citenamefont {Lal}, \citenamefont {Joe}, \citenamefont
  {Gan}, \citenamefont {Smadici}, \citenamefont {Finkelstein}, \citenamefont
  {Feng}, \citenamefont {Rusydi}, \citenamefont {Goldbart} \emph
  {et~al.}}]{lee2012two}%
  \BibitemOpen
  \bibfield  {author} {\bibinfo {author} {\bibfnamefont {J.~C.}\ \bibnamefont
  {Lee}}, \bibinfo {author} {\bibfnamefont {S.}~\bibnamefont {Yuan}}, \bibinfo
  {author} {\bibfnamefont {S.}~\bibnamefont {Lal}}, \bibinfo {author}
  {\bibfnamefont {Y.~I.}\ \bibnamefont {Joe}}, \bibinfo {author} {\bibfnamefont
  {Y.}~\bibnamefont {Gan}}, \bibinfo {author} {\bibfnamefont {S.}~\bibnamefont
  {Smadici}}, \bibinfo {author} {\bibfnamefont {K.}~\bibnamefont
  {Finkelstein}}, \bibinfo {author} {\bibfnamefont {Y.}~\bibnamefont {Feng}},
  \bibinfo {author} {\bibfnamefont {A.}~\bibnamefont {Rusydi}}, \bibinfo
  {author} {\bibfnamefont {P.~M.}\ \bibnamefont {Goldbart}},  \emph {et~al.},\
  }\href {http://www.nature.com/nphys/journal/v8/n1/full/nphys2117.html}
  {\bibfield  {journal} {\bibinfo  {journal} {Nature Physics}\ }\textbf
  {\bibinfo {volume} {8}},\ \bibinfo {pages} {63} (\bibinfo {year}
  {2012})}\BibitemShut {NoStop}%
\end{thebibliography}%
\end{document}